\documentclass[structabstract]{aa}
\usepackage{graphicx}
\usepackage{txfonts}
\usepackage{natbib}

\begin{document}

   \title{Polarized radiative transfer modeling of warped\\ and clumpy dusty tori}

   \titlerunning{Polarization from warped and clumpy dusty tori}

   \author{F.~Marin\inst{1}\thanks{\email{frederic.marin@astro.unistra.fr}}
   \and M.~Schartmann\inst{2,}\inst{3,}\inst{4}}
   
   \institute{Universit\'e de Strasbourg, CNRS, Observatoire astronomique de Strasbourg, UMR 7550, F-67000 Strasbourg, France
   \and Centre for Astrophysics and Supercomputing, Swinburne University of Technology, P. O. Box 218, Hawthorn, Victoria 3122, Australia
   \and Universit{\"a}ts-Sternwarte Mu{\"u}nchen, Scheinerstra{\ss}e 1, D-81679 M{\"u}nchen, Germany
   \and Max-Planck-Institut f{\"u}r extraterrestrische Physik, Postfach 1312, Giessenbachstr., D-85741 Garching, Germany}

   \date{Received May 18, 2017; Accepted August 21, 2017}

  \abstract
  {Active galactic nuclei (AGN) are anisotropic objects surrounded by an optically thick equatorial 
  medium whose true geometry still defy observers.}
  {In this paper, we aim to explore the optical, scattering-induced, polarization emerging from 
  clumpy and warped dusty tori to check whether they can fit with the unified model predictions.}
  {We run polarized radiative transfer simulations in a set of warped and non warped clumpy tori 
  to explore the differences induced by distorted dust distributions. We then include warped tori
  in a more complex model representative of an AGN to check, using polarimetry and imaging methods, 
  if warps can reproduce the expected polarization dichotomy between Seyfert-1 and Seyfert-2 AGN.}
  {The main results from our simulations highlight that isolated warped structures imprint the polarization 
  degree and angle with distinctive signatures at Seyfert-1 orientations. Included in an AGN 
  model, the signatures of warps are easily (but not always) washed out by multiple scattering in a 
  clumpy environment. Imaging polarimetry may help to detect warped tori but we prove that warps can 
  exist in AGN circumnuclear regions without 
  contradicting observations.}
  {Two warped tori with a non significant difference in geometry in terms of photometry or spectroscopy 
  can have totally different signatures in polarimetry. Testing the geometry of any alternative model 
  to the usual dusty torus using polarized radiative transfer is a necessary approach to verify or
  reject a hypothesis.}

\keywords{Galaxies: active -- Galaxies: Seyfert -- Polarization -- Radiative transfer -- Scattering}

\maketitle

\section{Introduction}
\label{Introduction}
It is now commonly accepted that the innermost regions of active galactic nuclei (AGN) are surrounded by a parsec-scale, optically thick,
equatorial distribution of dust. \citet{Antonucci1985} found, in the polarized optical light of the Seyfert-2 galaxy NGC~1068, a 
Seyfert-1 spectrum with broad emission lines, indicating that a Seyfert-1 nucleus exists in all Seyfert-2 AGN. The observational characteristics 
of Seyfert-2s are in fact hidden from direct view by optically thick material. Radiation form the central engine is scattered onto 
the AGN polar outflows, resulting in the observed near-ultraviolet, optical and near-infrared polarization properties. The idea of a common 
origin for emission lines for Seyfert-1 and 2 galaxies, broad- and narrow-line radio galaxies, and quasars is supported by the work of 
\citet{Shuder1981}. He found that H$\alpha$ and other optical emission lines appear to correlate over a range of 7 orders of magnitude with 
the continuum radiation. Subsequent recombinations that affect the emission lines might be uniform for almost all classes of AGN, but the
presence of this hardly-resolved dusty component is responsible for the observed orientation-dependent properties \citep{Antonucci1993,Urry1995}. 
Fortunately, the outskirts of the equatorial dusty medium of AGN start to be revealed thanks to the highest possible spatial resolution offered
by mid-infrared interferometry. \citet{Jaffe2004} spatially resolved a parsec-sized torus-shaped distribution of dust grains in the galaxy 
NGC~1068, where a small hot structure is embedded in a colder (320~K) dusty cocoon extending up to 3.4~pc in diameter. Further 
flux-limited interferometric studies of nearby quasars, using the MID-infrared Interferometric instrument (MIDI) at the Very Large Telescope 
Interferometer (VLTI; see, e.g., \citealt{Tristram2007,Burtscher2009,Burtscher2013,Tristram2014}) or the Keck interferometer 
\citep{Swain2003,Kishimoto2009a}, definitively confirmed the existence of the dusty torus albeit with large differences within the sample.

The morphology of this region is a poorly-constrained parameter of this multifaceted problem. The formation and hydrodynamic stability of a 
parsec-scale dust reservoir are not trivially explained by a homogeneous distribution of dust in a uniform toroidal arrangement. 
Detailed spatial studies revealed subtle problems with regards to smooth tori. Using Spitzer data, \citet{Sturm2006}, \citet{Hao2007}, 
\citet{Mason2009} and \citet{Nikutta2009} investigated the 10~$\mu$m silicate feature in a large sample of AGN. They found that the occasional 
detection of the feature in emission in Seyfert-2 AGN, together with the absence of any deeply absorbed features in Seyfert-1s 
might rule out smooth density torus models. To suppress the emission feature clumpiness is invoked 
\citep{Nenkova2002,Hoenig2006,Nenkova2008a,Nenkova2008b,Schartmann2008,Stalevski2012}. 
It is also a simple and elegant way to allow for the dust to survive the expected local temperatures. \citet{Krolik1988} postulated that the 
circumnuclear dusty material is likely clumpy, filled with a large number of individually optically thick clouds. Cloud merging and tidal 
shearing ensure a mixed environment that efficiently blocks radiation along the equatorial plane but allows a distant polar observer to see 
the central supermassive black hole (SMBH) through the dust-free funnel of the torus. Unfortunately, the exact geometry of the clumpy 
medium is beyond the resolution capabilities of current instruments. Infrared radiative transfer modeling has proven that multi-temperature, 
multi-phase clumpy distributions can reproduce most of the observed characteristics of nearby AGN \citep{Nenkova2008a,Nenkova2008b,Stalevski2012}. 
Similar results for the use of clumpy media instead of smooth density structures also emerged from polarized radiative transfer simulations 
\citep{Stalevski2012}. The near-infrared, optical and ultraviolet polarization signatures of a complex AGN model including a clumpy 
dusty torus were found to be in agreement with the predictions from the unified model \citep{Marin2015,Marin2015b}. 

If clumpiness is becoming more and more regarded as an important feature to add in (torus) simulations, there is another aspect of the 
unified model of Seyfert galaxies and quasars that is still barely investigated: the existence of warps. Using Very Long Baseline 
Array (VLBA) milliarcsecond-scale resolution observations of NGC~1068, \citet{Gallimore2004} revealed a slight misalignment between the line 
traced by the H$_2$O maser spots and the radio axis of the radio continuum source, believed to mark the location of the hidden nucleus. Very 
long baseline radio interferometry was also used to show that the accretion disk inside NGC~4258 may be a thin, subparsec-scale, differentially 
rotating warped disk \citep{Herrnstein1999}. 

Misaligned or warped structures have been detected in a variety of sources and can be imaged with great precision for disks around nearby 
stars, such as in the case of HD~142527. \citet{Marino2015} used polarized differential imaging to detect a warp in the outer disk of the HD~142527 
system, thanks to the shadows cast by the inner disk. Secular perturbations of a planet can also affect the inclinations (i.e. orbital plane) of 
nearby planetesimals. Introducing a planet into the disk on an orbit inclined to the disk midplane causes a warp to propagate away from the planet 
\citep{Augereau2001}. In the case of galaxies, \citet{Caproni2004a} and \citet{Caproni2004b} reported observational evidence of spin-induced 
disk/jet precession. This precession may lead to a warp of the innermost part of the AGN accretion disk with respect to the outer parts due to 
the Bardeen-Petterson effect \citep{Bardeen1975}. Warps can also be created at larger distances from the potential well by nonaxisymmetric 
magnetorotational instability \citep{Menou2001} as well as the Kelvin-Helmholtz instability \citep{Gunn1979}. The presence 
of shear motion inside the torus works as an agent to drive Kelvin-Helmholtz instabilities, resulting in the generation of a strong wind and 
potential warps \citep{Gunn1979,Kiuchi2015}. Such warps would be difficult to detect in photometry, spectroscopy or imaging in galaxies that 
are not nearly edge-on \citep{Meyer1989}, but polarimetry could be the answer. 

The discovery of a polarimetric dichotomy between Seyfert-1 and Seyfert-2 was the most important proof for the establishment of the unified 
scheme \citep{Antonucci1993}. Optical polarization from Seyfert-1s is mostly $<$~1\% and parallel to the projected radio axis of the system 
(its polarization position angle $PPA$ is thus equal to 90$^\circ$), while Seyfert-2s are always showing a polarization degree $>$~10\% with 
polarization angle = 0$^\circ$ (perpendicular polarization). This dichotomy is due to the presence of the circumnuclear dust material
that prevents direct radiation to escape from the equatorial plane in edge-on systems. Scattering inside the polar outflows results 
in higher polarization degrees $P$ and a perpendicular polarization, driven by the angle-dependence of Thomson scattering. This 
observational \citep{Antonucci1984} and theoretical \citep{Wolf1999,Young2000} model works very well for smooth density and clumpy 
tori \citep{Marin2012,Marin2015} but would warped dusty tori fit in this picture? Could they account for the high perpendicular
polarization found in Seyfert-2s? And would they reproduce the observed polarization dichotomy? The latest papers on warped structures 
around the innermost regions of AGN do not answer these questions and it is the goal of our present work. 

Accounting for both non-smooth density distributions and warps, we aim to test if warped media can affect the optical continuum 
polarization of a complex AGN model. We want to determine if warps and misaligned structures at the torus distances can fit the unified 
model, such as postulated by \citet{Lawrence2010}, or if caution must be taken when dealing with deformed tori. To do so, we construct 
a simple toy model for warped tori in Sect.~\ref{Modeling}, where we present the formalism and the Monte Carlo radiative transfer code 
used in our simulations. It is not our ambition to provide an accurate physical model for warped structures; this should be done together 
with hydrodynamical simulations \citep[e.~g.~][]{Wada2016}. In Sect.~\ref{Isolated}, we test our clumpy and warped torus and compare the 
results to non-warped structures. We explore the space of parameters used to create the warp and show that warps have a profound impact 
on isolated tori. We include the warped structures in more complex AGN models in Sect.~\ref{AGN}, taking into account polar outflows 
and equatorial inflows. We create imaging polarimetry maps of an AGN with a warped torus before exploring another model 
for a warped, uniform, dusty torus in Sect.~\ref{Jud}. We discuss our results and further progress to be made in Sect.~\ref{Discussion} 
before concluding in Sect.~\ref{Conclusion}.

\section{Modeling warped and clumpy dusty tori}
\label{Modeling}
One of the most common geometries used to represent the circumnuclear dusty material in AGN is a torus (see the earliest representations 
in \citealt{Antonucci1984}, where the outer boundary of the equatorial region was not drawn due to the lack of information on that point). 
Other morphologies, such as flared dusty disks, representative of outflowing material, are sometimes used in infrared studies (e.g. 
\citealt{Manske1998,Stalevski2012}) but give similar spectroscopic results to dusty tori\footnote{For a polarimetric investigation of the 
differences between a clumpy flared disk and a clumpy torus, see \citet{Marin2015b}.}. In this paper, we focus on a toroidal structure 
that will be warped at a given distance from the center of the model (where the SMBH lies).

\subsection{The clump distribution}
\label{Modeling:Distribution}

\begin{figure*}[ht] 
   \begin{minipage}[c]{.5\linewidth}
      \centering
      \includegraphics[trim = 0mm 17mm 0mm 17mm, clip, width=0.75\linewidth]{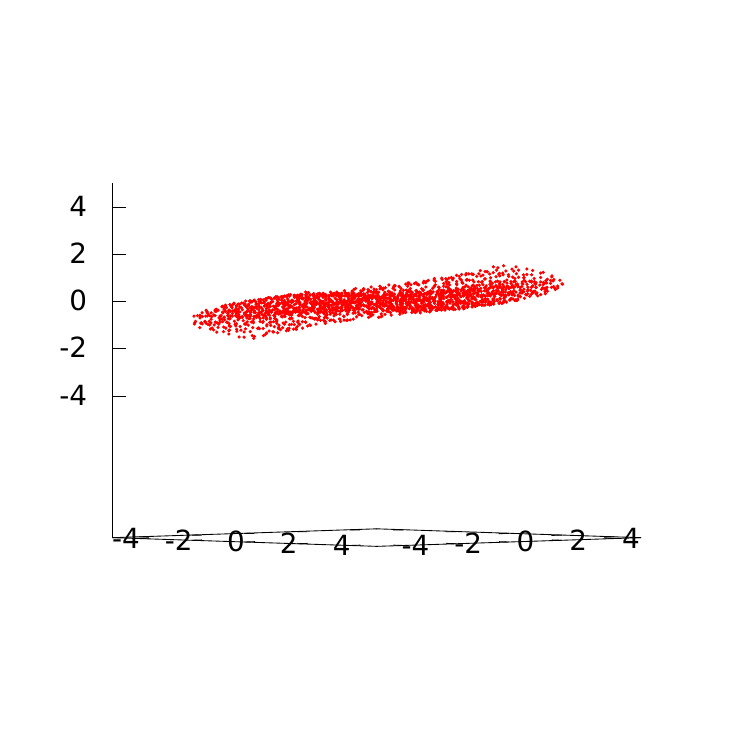}
      \caption{(a) 3-dimensional view.} 
      \vspace{4ex}
   \end{minipage} \hfill
   \begin{minipage}[c]{.5\linewidth}
      \centering   
      \includegraphics[width=0.5\linewidth]{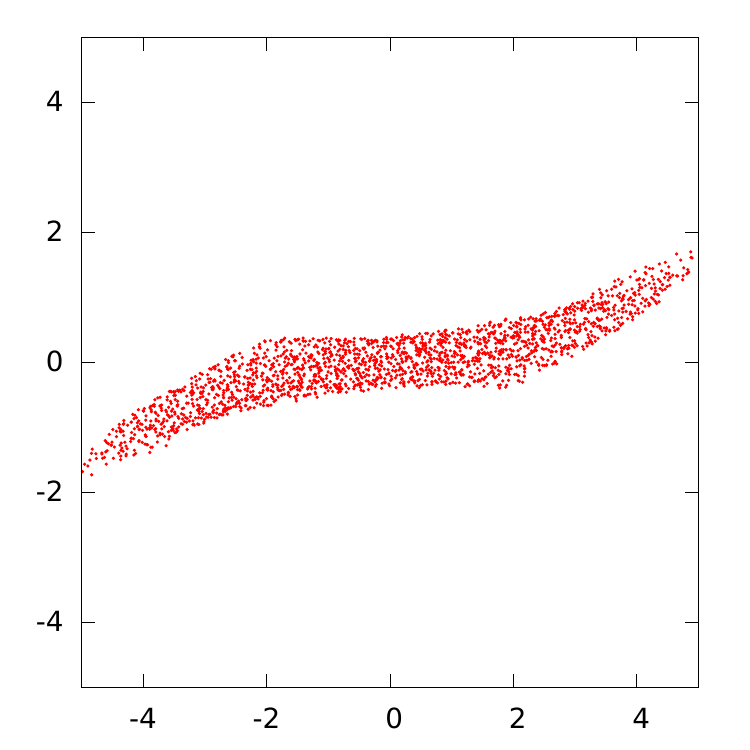}
      \caption{(b) Projection onto the $xz$-plane (Seyfert-2 view, azimuthal angle = -90$^\circ$).} 
      \vspace{4ex}
   \end{minipage}
   \begin{minipage}[c]{.5\linewidth}
      \centering     
      \includegraphics[width=0.5\linewidth]{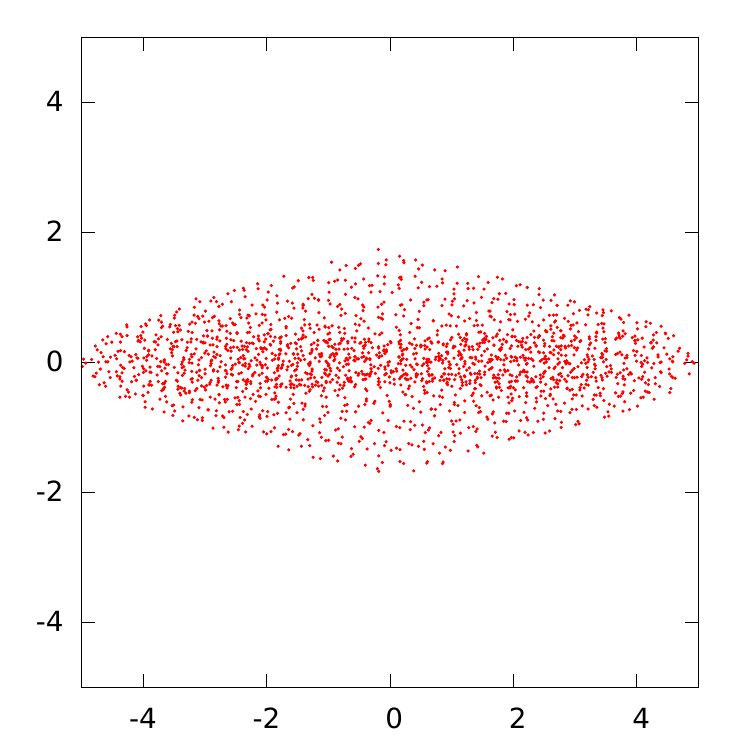}
      \caption{(c) Projection onto the $yz$-plane (Seyfert-2 view, azimuthal angle = 0$^\circ$).} 
   \end{minipage} \hfill
   \begin{minipage}[c]{.5\linewidth}
      \centering     
      \includegraphics[width=0.5\linewidth]{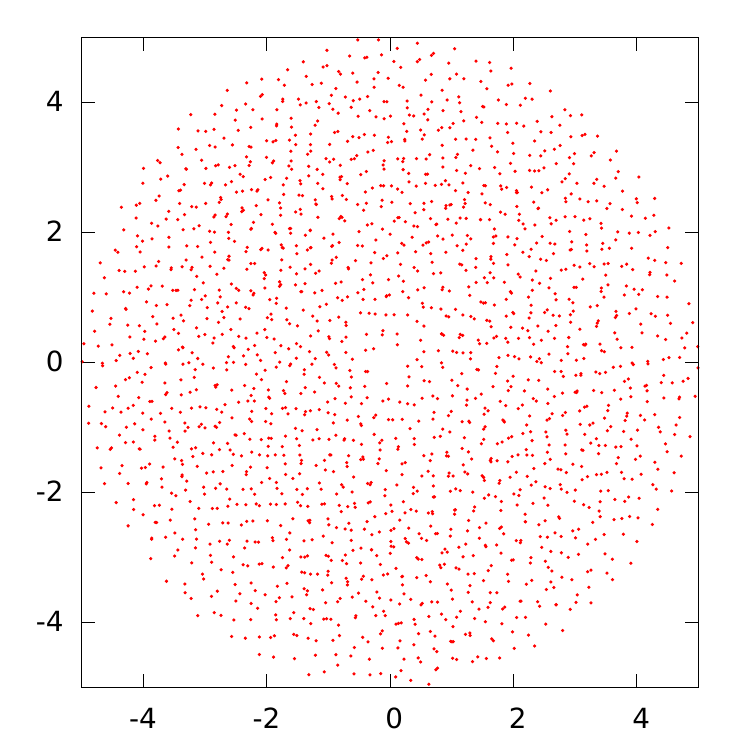}
      \caption{(d) Projection onto the $xy$-plane (Seyfert-1 view).} 
   \end{minipage}   
   \setcounter{figure}{0}
   \caption{Representations of a clumpy, warped, dusty torus 
	  with R$_{\rm warp}$ = 2~pc and $\theta_{\rm warp}$ 
	  = 30$^\circ$. Axes are in parsecs.}
  \label{Fig:3Dplots} 
\end{figure*}

We first consider a regular torus with an elliptical cross-section, ranging from an inner radius R$_{\rm in}$ = 0.1~pc to an outer 
radius R$_{\rm out}$ = 5~pc. The inner radius is fixed according to the dust sublimation radius for a $\sim$10$^7$~M$_\odot$ SMBH 
with sub-Eddington accretion rate. R$_{\rm sub}$ is correlated to the UV/optical continuum emission (L$_{\rm bol}$/L$_{\rm Edd}$ = 0.45, 
see \citealt{Kollmeier2006}) and to the sublimation temperature T$_{\rm sub}$. The latter is fixed to 1500~K, a value that accounts 
for the temperature range of ambient gas pressures for both graphite and silicate grains \citep{Kishimoto2007}. The outer torus radius 
corresponds to the typical extent of compact and optically thick tori in nearby AGN, ranging from 1 to 10~pc according to simulations 
\citep{Pier1992,Siebenmorgen2015} and observations \citep{Tristram2007,Kishimoto2009}. The torus half-opening angle is fixed to 30$^\circ$ 
from the equatorial plane, a value chosen to agree with observed half-opening angles of the equatorial AGN region recently estimated by 
\citet{Sazonov2015} and \citet{Marin2016}. The toroidal structure is then filled with constant-density spheres with radius 0.11~pc
\citep{Stalevski2012}. Each clump has an optical depth of 50 in the V-band, so a single cloud can efficiently obscure 
the central AGN engine in the optical band \citep{Siebenmorgen2015}. The dust mixture has a composition similar to the Milky Way dust 
\citep{Mathis1977}. Note that the choice of a Milky Way dust mixture instead of, e.g., a Small Magellanic Cloud like dust composition, 
where the dust is smaller than that in the Galactic diffuse ISM \citep{Prevot1984}, does not change the outcomes of the simulations in 
the optical band as scattering in opaque dust clouds produces relatively grey scattering \citep{Kishimoto2001}. We fix the number of 
clouds to 2000 in order to have a volume filling factor of 22--23\% (the spherical clouds do not overlap, see also 
\citealt{Stalevski2012}). This model is the basis of all models we will explore in the forthcoming sections, and it will be studied 
in Sect.~\ref{Modeling:Benchmark} as a test case. 

To create warps in this structure, we define a radius R$_{\rm warp}$ at which the clumps will depart from the torus structure. 
When a clump is generated at a distance larger than R$_{\rm warp}$, its $z$-coordinate will be shifted upward or downward 
according to its position on the $xy$-plane:

\begin{equation}
  z = \pm [z + (\sqrt{x^2+y^2}-R_{\rm warp})\tan(\theta)] \,,
  \label{eq1}
\end{equation}

with

\begin{equation}
  \theta = \frac{2\times\theta_{\rm warp}(\frac{\pi}{2}-\arctan(\frac{|y|}{|x|}))}{\pi}.
  \label{eq2}
\end{equation}

$\theta_{\rm warp}$ is the maximum warping angle, i.e. the angle between the equatorial plane and the warped torus in the $xz$-plane.
A torus warped through angle $\theta_{\rm warp}$ will intercept and reradiate a fraction close to $\theta_{\rm warp}$/3 of the luminosity 
of the central source \citep{Phinney1989}. A visual representation of a warped torus with R$_{\rm warp}$ = 2~pc and $\theta_{\rm warp}$ = 30$^\circ$ 
is shown in Fig.\ref{Fig:3Dplots}. The top-left panel is a three-dimensional view of the model with 2000 clumps (clump sizes not to scale).
The other three panels show the projection of the warped torus structure onto the $xz$, $yz$, and $xy$-planes. The distance at which the 
warping effect starts can be seen in Fig.\ref{Fig:3Dplots}~(b), together with the warping angle $\theta_{\rm warp}$. On 
Fig.\ref{Fig:3Dplots}~(c), one can discern a denser region close to the equator, which is the inner part of the torus that is unaffected 
by the warping effect. Finally, panel (d) is a view from the top and it is impossible to detect the warp.

This simple warping method has several advantages: the outer radius of the dusty torus is the same regardless of the azimuthal angle of 
the observer, so Seyfert-1s with low inclinations ($<$ 10$^\circ$) should not lead to excessively elongated dust lanes along any directions. 
Only the height of the structure is affected, resulting in possible polar dust signatures for intermediate and Seyfert-2s, way beyond 
the torus height defined by its half-opening angle \citep{Bock2000,Burtscher2013}. Polar mid-IR emission from this kind of warped dust distribution 
could contribute to the important polar infrared flux discovered by interferometry studies of Seyfert-2s with high position angle and baseline 
coverage \citep{Tristram2014,Asmus2016,Lopez2016}. The antisymmetry between the upper and lower parts of the warped structure 
naturally arises even under these simple conditions. This could be due to a strong radiation field or an outflow dense enough to wipe the 
external surface of the circumnuclear dust, but they must sustain a small deflection with respect to the torus axis. This is the basic idea 
developed in \citet{Pringle1996}, \citet{Nayakshin2005}, and \citet{Lawrence2010}, see Fig.~5 and 6 in \citet{Lawrence2010} for illustrations. 

There are other methods to warp tori. Recently, \citet{Jud2017} investigated uniform, geometrically-thin, warped dusty disks in the infrared 
band using a disk modified to have a concave shape at small radii which turns over at large radii. Their final structure is similar to 
what is being investigated here, with differences in the temperature profile and bulk shape of the disk at R $>$ R$_{\rm warp}$. Since 
we will explore the optical, scattering-induced, linear polarization of warped tori in this paper, the temperature profile has very little 
influence onto the scattered radiation. An important difference is that our modeling includes clumpiness, a critical aspect that would change the 
shape of the silicate feature and the spectral energy distribution in the infrared band (\citealt{Jud2017} and references therein). However, note 
that our model was not checked for consistency with the IR emission expected from Seyferts, e.g. the silicate feature strengths or interferometry 
results. Our goal is not to present a new warping method but to check whether warped tori can exist within the unified AGN picture, based on 
polarimetric observations.

\subsection{Radiative transfer with STOKES}
\label{Modeling:STOKES}
Radiative transfer was achieved using the Monte Carlo code {\sc stokes}, a simulation tool that was written to investigate the polarization 
signature of AGN in the optical and ultraviolet bands \citep{Goosmann2007}. The code was later improved to provide imaging analysis \citep{Marin2012}
and clumpy reprocessing regions \citep{Marin2015}. A summary of the code performances can be found in \citet{Marin2014}. The three-dimensional
capabilities of the code allow to explore an AGN model without any symmetries by virtually revolving around. For the remainder of this paper,
we will investigate three different inclinations: 18$^\circ$ (Seyfert-1 view), 50$^\circ$ (intermediate inclination), and 87$^\circ$ 
(Seyfert-2 view). 

The photon source used in the simulation is a point-like source of isotropic emission. The initial radiation is unpolarized and fixed at a 
monochromatic wavelength of 5000~\AA (B/V-band). Around the central source are the reprocessing regions that can scatter, absorb or re-emit radiation.
Mie and Thomson scattering are accounted for but the code does not handle temperature variation due to dust being heated or grain cooling by emission at 
infrared wavelengths. Re-emission from dust is thus not treated in theses simulations, which is a non-vital simplification as dust re-emits at much longer 
wavelengths than in the blue and visual bands. The four Stokes parameters of light \citep{Stokes1851} are recorded but we will focus on linear 
polarization, since optical circular polarization from AGN was much less investigated by observers. As the Monte Carlo method employs a stochastic 
approach, the results obey Poisson statistics. We set the number of photons to be simulated in order to have small ($<$~1\%) statistical fluctuations
per direction bin. Hence, each simulation sampled $\sim$ 10$^9$ photons for a computational time of 96~hours per model\footnote{Radiative transfer 
simulations are time-consuming, especially when clumpy media are involved. After emission or a scattering event, {\sc stokes} checks if a clump is along 
its direction. As there are thousands of clumps in the model, the checking procedures drastically increases the duration of the simulations.}.

\subsection{Test case: a clumpy dusty torus without warps}
\label{Modeling:Benchmark}

\begin{figure}
   \centering
   \includegraphics[trim = 0mm 0mm 0mm 0mm, clip, width=9cm]{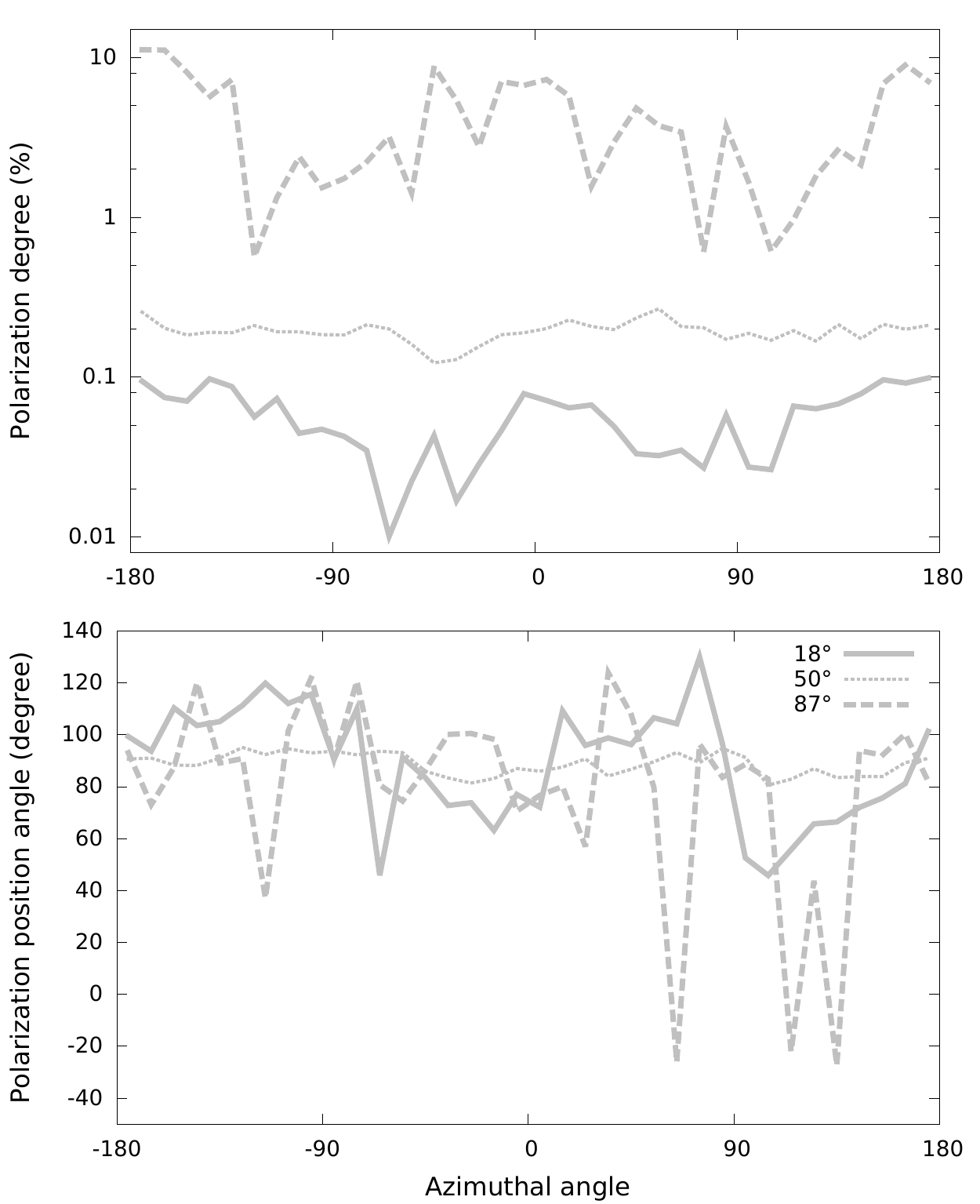}
   \caption{Optical polarization (5000~\AA) from a clumpy dusty 
	    torus without warping effects. The top panel shows 
	    the polarization degree $P$ and the bottom 
	    panel the polarization position angle $PPA$.
	    The lines refer to the inclination of the model:
	    the solid line indicates a viewing angle of 
	    18$^\circ$ (Seyfert-1 view), the dotted line 
	    an intermediate inclination of 50$^\circ$,
	    and the long-dashed line a viewing angle of 
	    87$^\circ$ (Seyfert-2 view).}
  \label{Fig:Benchmark}
\end{figure}

Before exploring the optical, scattering-induced, linear polarization from warped and clumpy dusty tori, the test model presented in 
Sect.~\ref{Modeling:Distribution} was investigated. This test case is a clumpy torus model without warp and is similar to what has been studied in 
\citet{Marin2015}. The effects of the clumpiness of the torus in a direct comparison with a continuous torus are already presented in \citet{Marin2015} 
and \citet{Marin2015b}. The results are shown in Fig.~\ref{Fig:Benchmark}: the top graph presents the polarization degree $P$ as a function of the azimuthal 
angle for three different inclinations $i$. The chaotic variations in $P$ are not due to insufficient Monte Carlo statistics. They are due to the clumpiness 
of the circumnuclear region: as the circumnuclear torus was randomly filled with clumps, each azimuthal angle sees the central source with a different covering 
factor. The photon trajectories and the absorption probabilities vary from two consecutive bins separated by an angle of 1 degree, resulting in the 
observed variations in $P$. The linear polarization is also inclination-dependent, with $P <$ 1\% for Seyfert-1 inclinations and $P <$ 10\% at Seyfert-2 inclinations. 
Compared to the pole-on and edge-on views, the intermediate inclination polarization curve appears smoother in azimuth dependence; this effect is due to a lesser 
influence of clumpiness onto the $i$ = 50$^\circ$ case. At the transition angle between Seyfert-1s and Seyfert-2s, back scattering onto the torus funnel is contributing 
the most to the polarization signal: in this case, the photon only ``sees'' a convex wall of dust and, if not absorbed, backscatters towards the observers, 
regardless of the existence of clumps behind the first cloud. An azimuthal-integration of $P$ gives the following average values: 0.04\% at $i$ = 18$^\circ$, 
0.19\% at $i$ = 50$^\circ$, and 2.61\% at $i$ = 87$^\circ$. Since the variations of the non-warped model are caused by the random clump distributions, 
we will use those integrated values for comparison purposes with our following models (see next section).

Similar conclusions can be drawn from the polarization position angle ($PPA$, bottom figure), which is mainly parallel to the symmetry axis of the model
(hence yielding a value close to 90$^\circ$ when azimuthally-integrated). This is the averaged polarization angle already found in previous modeling 
with geometrically flat, clumpy, tori, see \citet{Marin2015b} and Fig.~5 in \citet{Marin2015}. What is shown here and not in the previous publications is that the $PPA$ 
can rotate by 90$^\circ$ when the covering fraction of the source is insufficient at Seyfert-2 inclinations. This effect is strengthened by the fact that there
is no diffuse inter-clump medium that would prevent or attenuate the propagation of photons in edge-on orientations. A second feature of interest for the 
remainder of the paper is that the $PPA$ can be offsetted by 0 to 30$^\circ$ from parallel polarization at a given phase, simply because of the random clump
distribution. 

The results from the test case seem to contradict the observations, where Seyfert-2s should have P $>$ 10\% and PPA $\sim$ 0$^\circ$, but we remind 
the reader that it is only because the torus is isolated from the rest of the usual AGN components. We will see in the following sections that including 
polar winds will drastically change the picture.

\section{Optical polarization from clumpy and warped tori}
\label{Isolated}

\subsection{Results for a specific parametrization}
\label{Isolated:Comparison}

\begin{figure}
   \centering
   \includegraphics[trim = 0mm 0mm 0mm 0mm, clip, width=9cm]{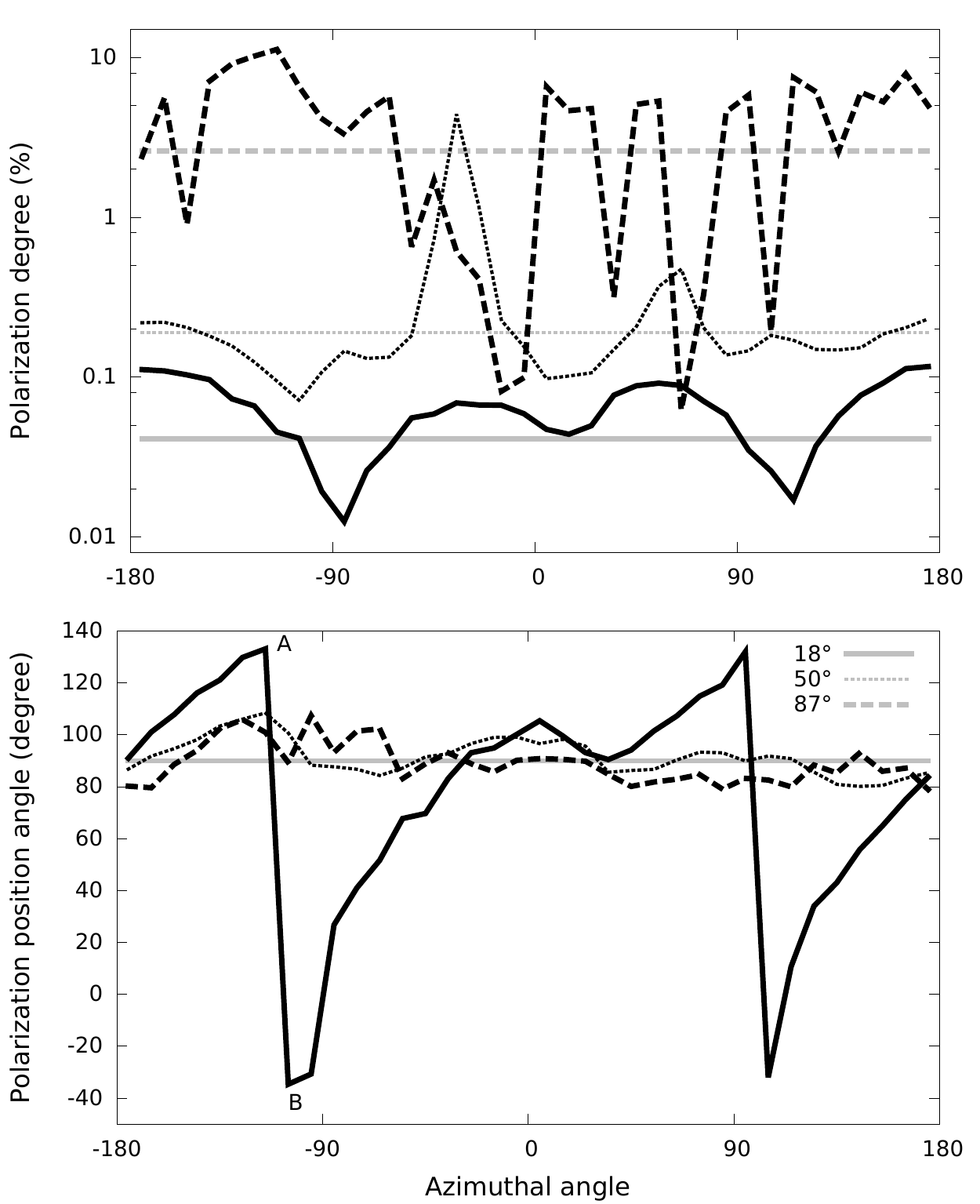}
   \caption{Optical polarization (5000~\AA) from a warped and clumpy 
	    dusty torus (in black) with R$_{\rm warp}$ = 2~pc and 
	    $\theta_{\rm warp}$ = 30$^\circ$, overplotted onto the 
	    azimuthally-integrated polarization from the test 
	    model (in gray). The top panel shows the polarization 
	    degree $P$ and the bottom panel the polarization position 
	    angle $PPA$. Points A and B mark the azimuthal 
	    bins before and after the large amplitude variation in $PPA$ 
	    discussed in the text and developed in Fig.~\ref{Fig:Rotation}.	    
	    The legend is the same as in 
	    Fig.~\ref{Fig:Benchmark}.}
  \label{Fig:Comparison}
\end{figure}

\begin{figure}
   \centering
   \includegraphics[trim = 0mm 0mm 0mm 0mm, clip, width=9cm]{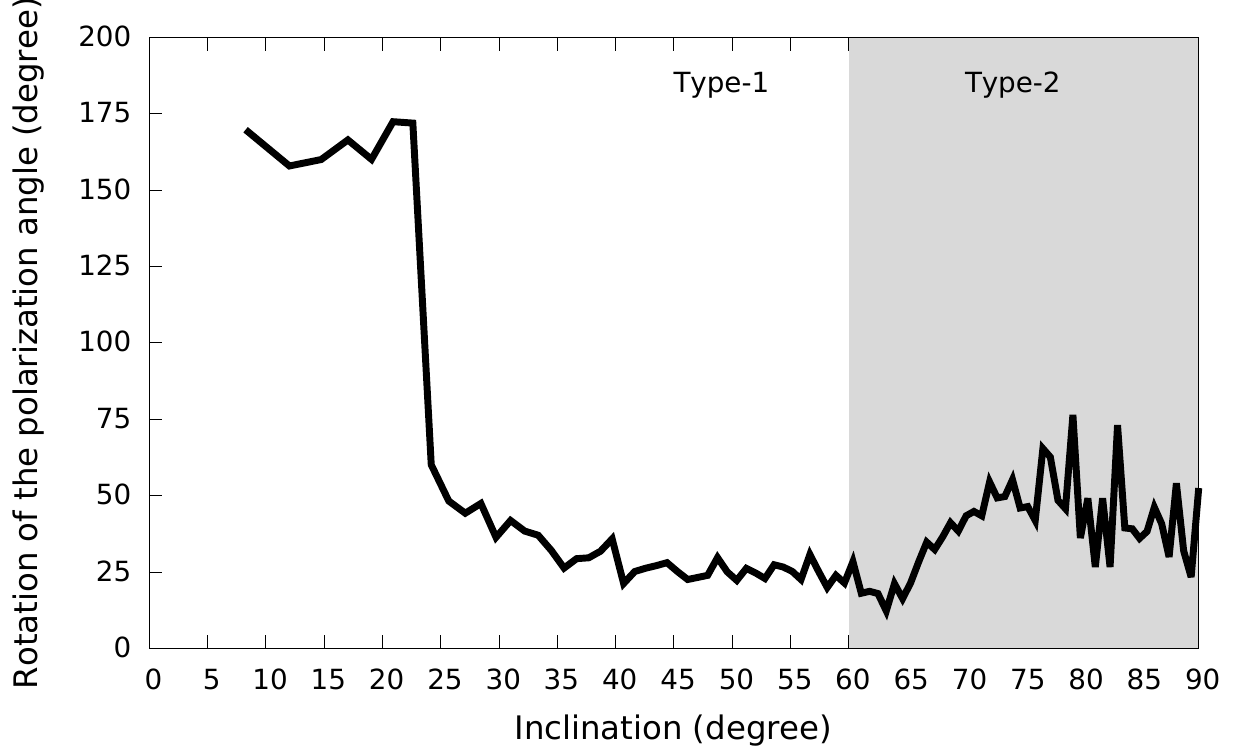}
   \caption{Rotation of the $PPA$ between points A and B 
	    (see Fig.~\ref{Fig:Comparison}), corresponding to 
	    the difference between maximum and minimum values 
	    of the $PPA$ at azimuthal angles close to -90$^\circ$. 
	    The polarization angle modulation is shown for every inclination 
	    of the observer. The shaded gray area represents the line-of-sights
	    obscured by the clumpy equatorial dust distribution,
	    setting the boundary between Seyfert-1s and Seyfert-2s
	    (the later type being in the gray area).}
  \label{Fig:Rotation}
\end{figure}

In the previous section, we obtained the optical, scattering-induced, polarimetric signature of a clumpy toroidal structure and saw that chaotic 
variations in $P$ and $PPA$ naturally arise from the random distribution of dust clumps that only fill 22-23\% of the total volume. We now add a warp that 
starts at a distance of 2~pc from the center of the model. The warp has an angle of 30$^\circ$ with respect to the equatorial plane, thus reaches a maximum 
height of 1.73~pc above the midplane (see Fig.~\ref{Fig:3Dplots}b). 

Compared to the test case, a warped clumpy torus (Fig.~\ref{Fig:Comparison}) shows remarkable signatures in both $P$ and $PPA$ at a Seyfert-1 
inclination. The $PPA$ presents sawtooth oscillations with sharp inversions around azimuthal angles -90$^\circ$ and +90$^\circ$, associated with 
local minima in $P$. The polarization angle modulation can be as high as 180$^\circ$ due to the maximum asymmetry seen by a distant observer\footnote{The 
polarization position angle is periodic. Any polarization ellipse is indistinguishable from one rotated by 180$^\circ$, which is one of the fundamental advantages of radio 
antenna polarimetry. The sawtooth pattern is caused by this periodicity; an observer is expected to see a continuous increase of the PPA with a polarization modulation.}.
The bottom part of the warp being hidden by the equatorial dust content at $i$ = 18$^\circ$, only the side of the warp that faces the observer 
contributes to the net polarization, resulting in azimuthal-dependent $PPA$. This effect is much less pronounced at intermediate and equatorial 
inclinations: the inner parts of the clumpy structure efficiently block radiation and the specific sawtooth oscillations are smoothed out. 
Compared to the test model, the scatter in $PPA$ for the warped torus appears smaller at equatorial inclinations. Polarization mainly arises
from backscattering from the torus funnel and the contribution of the warped and distant surfaces is negligible (as it will be confirmed by 
imaging polarimetry results, see Sect.~\ref{AGN:Maps}). At those inclinations, the averaged $P$ for a warped dusty torus is marginally 
higher than what has been found for a non-warped model. Even if $P$ shows more scatter around the mean for the warped model, we may estimate 
its averaged values: 0.06\% at $i$ = 18$^\circ$, 0.34\% at $i$ = 50$^\circ$, and 4.23\% at $i$ = 87$^\circ$.

We further investigated the sharp variation in $PPA$ by looking at the model from a wider range of inclinations. In Fig.~\ref{Fig:Rotation}, 
the rotation of the polarization angle close to the minimum of the sawtooth ($\sim$~90$^\circ$ phase, between points A and B in Fig.~\ref{Fig:Comparison}) 
is shown in black. The code has difficulties to sample inclinations lower than 8$^\circ$, but it is clear that the largest $PPA$ variations occur 
at polar inclinations. The increasing importance of backscattering from the torus funnel at inclinations $i \ge$ 25$^\circ$ stabilizes the polarization 
angle at low polarization modulations; the $PPA$ values are consistent with clump-induced effects, as discussed in the last point in Sect.~\ref{Modeling:Benchmark}. 
The inclination at which the $PPA$ flattens is consistent with $\theta_{\rm warp}$ = 30$^\circ$, accounting for the effects of clumping.
At $i$ = 60$^\circ$, torus obscuration starts but the gap between the clouds allows radiation to easily escape. At $i \ge$ 65$^\circ$, the $PPA$ 
starts again to oscillate due to the growing influence of equatorial obscuration. Only the fraction of the torus funnel that is opposite to the 
observer and the far end of the warped structure are scattering photon, but the resulting polarization angle is still dominated by multiple 
scattering effects. Thus, clumpiness has a much more important impact than warps on polarization at intermediate and edge-on orientations.

\subsection{Exploring the parameter space}
\label{Isolated:Tables}

\begin{figure*}
   \centering
   \includegraphics[trim = 0mm 0mm 0mm 0mm, clip, width=18cm]{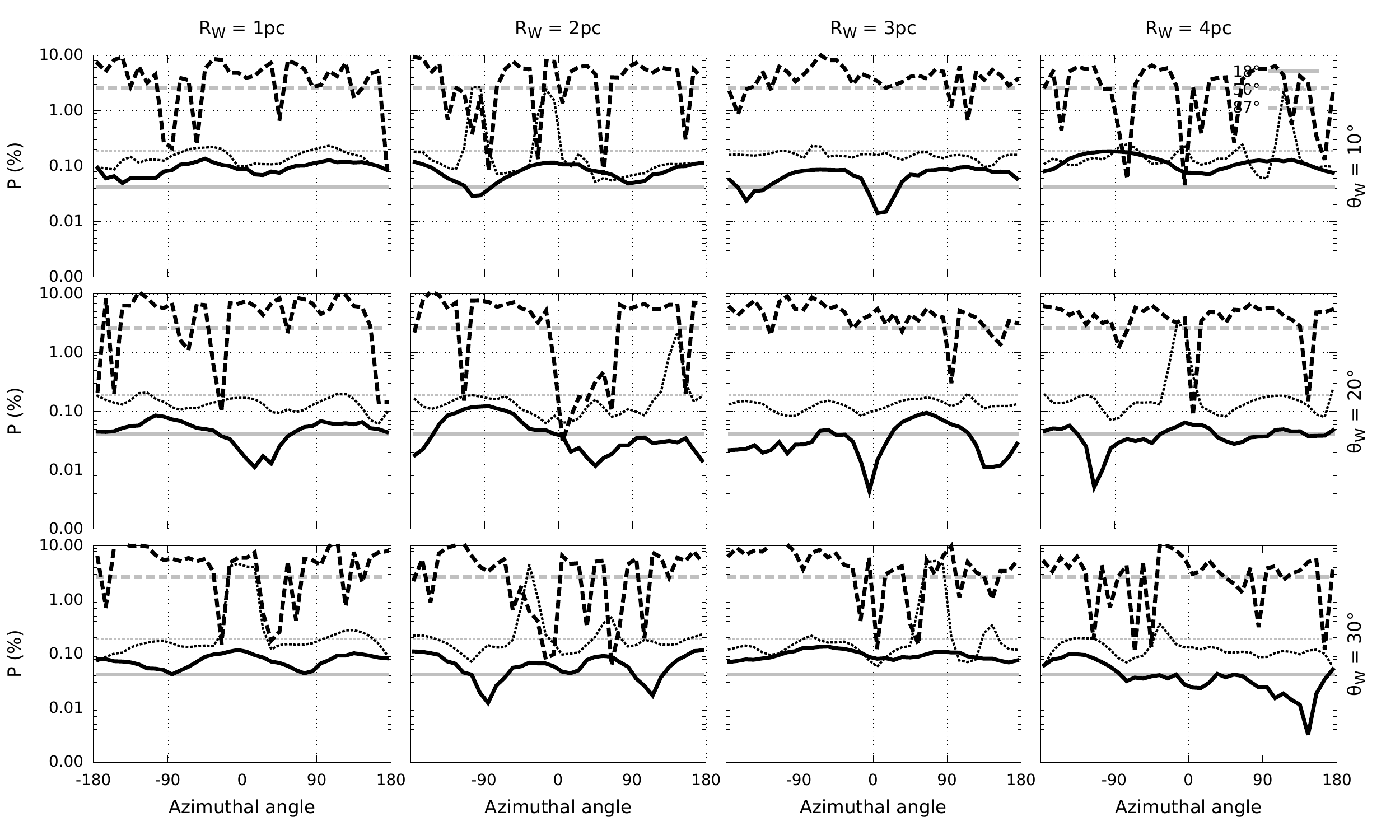}
   \caption{Resulting percentage of polarization $P$ as a function
           of viewing angle (solid line: 18$^\circ$, dotted line:
           50$^\circ$ and long-dashed line: 87$^\circ$) and 
           azimuthal angle for a set of warped and clumpy dusty 
           tori. The legend is the same as in Fig.~\ref{Fig:Benchmark}.}
  \label{Fig:Table_Isolated_PO}
\end{figure*}

\begin{figure*}
   \centering
   \includegraphics[trim = 0mm 0mm 0mm 0mm, clip, width=18cm]{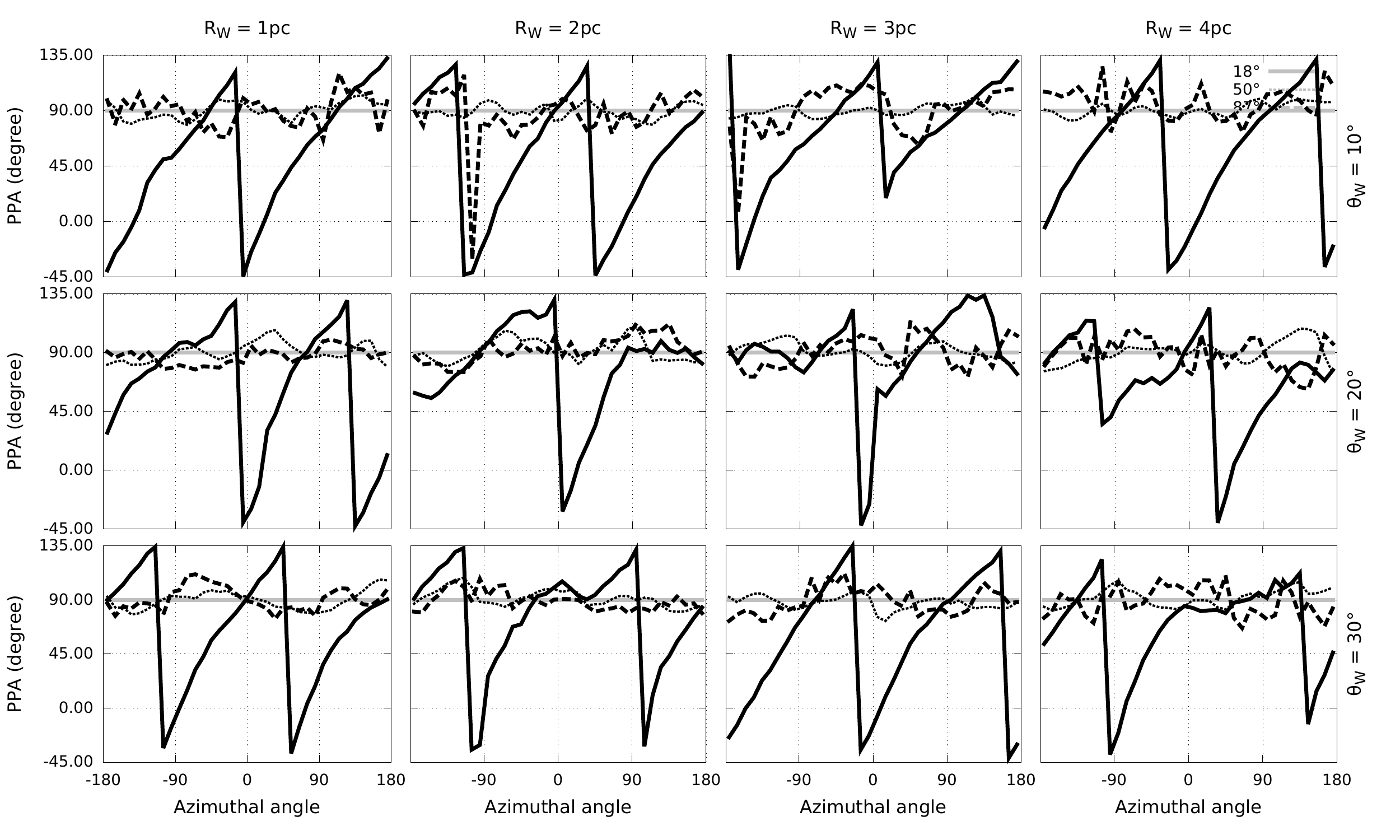}
   \caption{Same as Fig.~\ref{Fig:Table_Isolated_PO}, except that
	    the polarization position angle $PPA$ is shown here.}
  \label{Fig:Table_Isolated_PA}
\end{figure*}

There are two important parameters that drive the warp structure: the distance at which the warping effect starts (R$_{\rm warp}$) and the maximum
angle between the equatorial plane and the warped structure ($\theta_{\rm warp}$). Both can be used as tracers of the initial non-axisymmetric 
phenomenon that leads to torus warping but the question is: can they be estimated from polarimetric measurements? In this subsection, we run a grid of 
warped tori varying R$_{\rm warp}$ and $\theta_{\rm warp}$, and we examine the resulting polarization as a function of azimuthal angle. The results 
are summarized in Fig.~\ref{Fig:Table_Isolated_PO} and Fig.~\ref{Fig:Table_Isolated_PA} for the polarization degree and the polarization angle, 
respectively. 

The first and probably most important observation is that a warped dusty torus always drives 180$^\circ$ rotations of the polarization angle at 
Seyfert-1 orientations when revolving around the structure. The sawtooth polarization modulation in $PPA$ are associated with local minima in $P$, such as seen in 
Fig.~\ref{Fig:Comparison}, that can decrease $P$ by one order of magnitude in the most extreme case. The azimuthal angle at which the polarization modulation happens 
does not change gradually as a function of both R$_{\rm warp}$ and $\theta_{\rm warp}$; it cannot be easily predicted. The random sampling of the 
density distribution/clumpiness creates higher- and lower-density sub-regions, resulting in enhanced asymmetry, and partly offset from 90$^\circ$ in 
azimuthal angle the minima in $P$ and $PPA$. Regardless of the configuration of the warp, the sawtooth rotations of the $PPA$ is a clear and distinct 
signature of isolated warped tori. The averaged polarization degree for warped tori at Seyfert-1 inclinations ranges from 0.04\% to 0.12\%. These values 
are slightly higher than what was found for non-warped, non-clumpy torus models \citep{Goosmann2007}, as asymmetry is enhanced here. At other inclinations, 
$PPA$ flattens around 90$^\circ$ due to the randomization of the clump positions and multiple scattering that tend to hide potential 
$P$(R$_{\rm warp}$,$\theta_{\rm warp}$) or $PPA$(R$_{\rm warp}$,$\theta_{\rm warp}$) correlations. Variations in the covering fraction of the source 
at different azimuthal angles have a much deeper effect onto the continuum polarization than warps. The averaged polarization degree for warped tori 
at intermediate inclinations ranges from 0.13\% to 0.62\%; for edge-on systems, $P$ ranges from 3.41\% to 5.61\%, which is in agreement with past 
polarimetric modeling \citep{Marin2015}.

\section{Impact of warped dusty tori in a global AGN model}
\label{AGN}
We can summarize our current results by stating that warps do not strongly affect the polarimetric signatures of clumpy dusty tori at intermediate 
and edge-on inclinations. The signatures of warps in the dust structure are washed out by the polarization emerging from reprocessing 
on the equatorial distribution of clumps. However, at Seyfert-1 inclinations, there is a tremendous impact of warps onto the polarization position angle.
This 180$^\circ$ polarization modulation occurs in the case of isolated warped structures, but does this feature also impact a complex AGN model where inflowing
material from the torus connects to the broad line region (BLR)? Where ionized polar outflows extend up to tens of parsecs? Moreover, can warped and clumpy dusty 
tori account for the high perpendicular polarization found in Seyfert-2s? It is the aim of the next section to answer those fundamental questions.

\subsection{Polarization from an AGN with a warped equatorial obscurer}
\label{AGN:Tables}

\begin{figure*}
   \centering
   \includegraphics[trim = 0mm 0mm 0mm 0mm, clip, width=18cm]{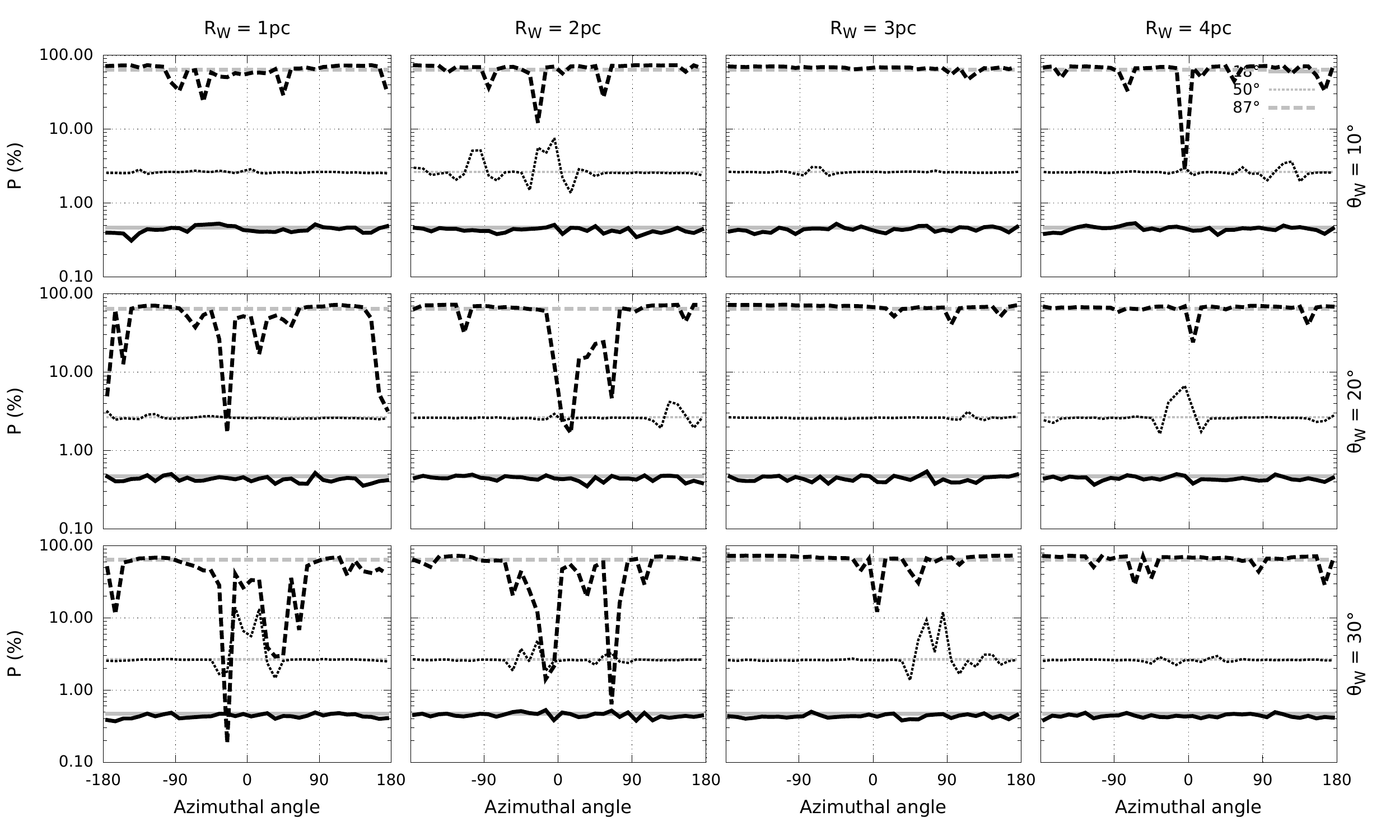}
   \caption{Resulting percentage of polarization $P$ as a function
           of viewing angle (solid line: 18$^\circ$, dotted 
           line: 50$^\circ$ and long-dashed line: 87$^\circ$) 
           and azimuthal angle for a set of complex AGN models 
           (see text) using the same warped and clumpy dusty tori 
           as in Fig.~\ref{Fig:Table_Isolated_PO}. Both the black 
           and gray lines describe the complex AGN model, but the 
           gray line is the result of an azimuthal integration 
           (fine details are thus washed out). The rest of       
           the legend is the same as in Fig.~\ref{Fig:Benchmark}.}      
  \label{Fig:Table_AGN_PO}
\end{figure*}

\begin{figure*}
   \centering
   \includegraphics[trim = 0mm 0mm 0mm 0mm, clip, width=18cm]{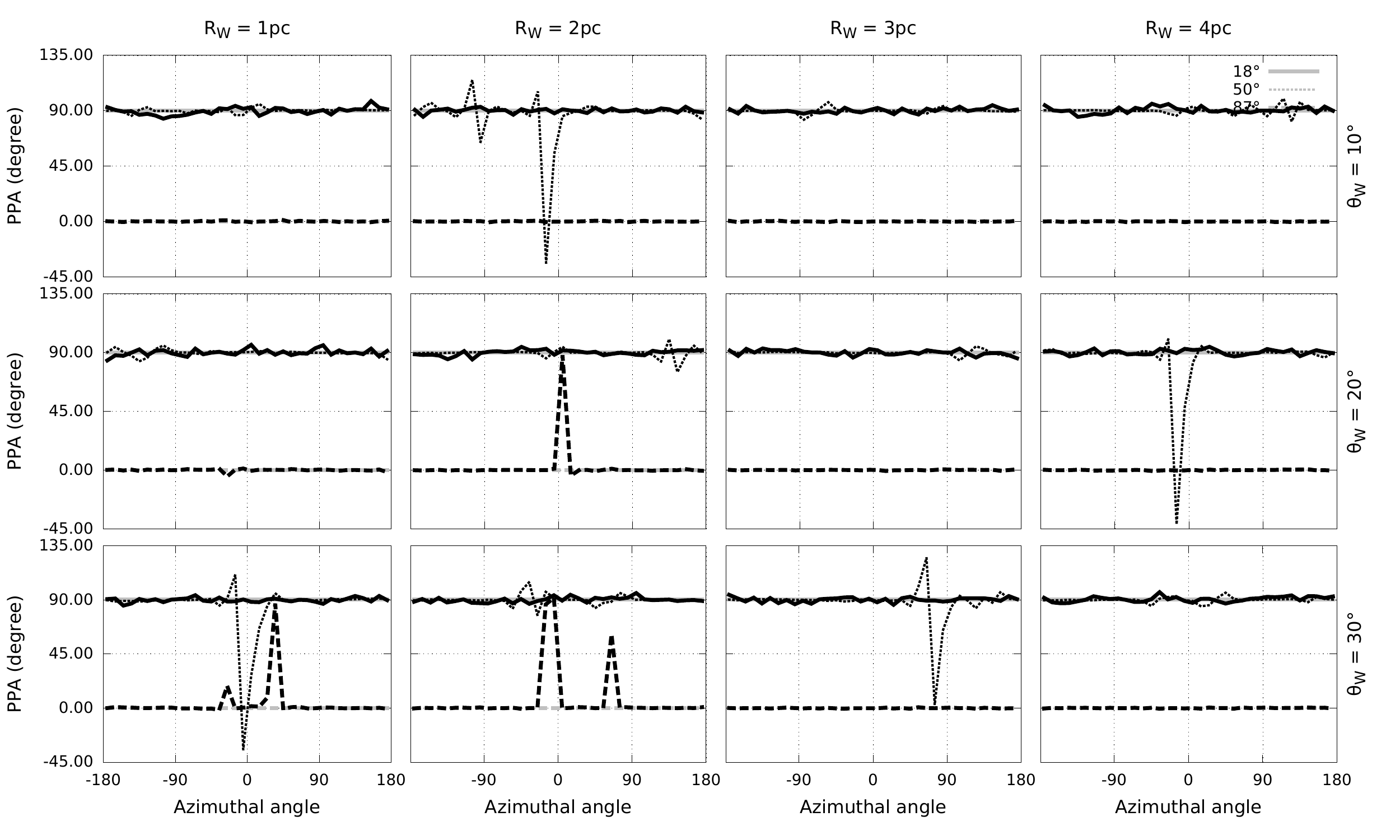}
   \caption{Same as Fig.~\ref{Fig:Table_AGN_PO}, except that
	    the polarization position angle $PPA$ is shown here.}
  \label{Fig:Table_AGN_PA}
\end{figure*}

To investigate a complex AGN model, we added to the previous warped and clumpy dusty tori two extra components: an equatorial, electron-dominated
inflow and a polar, bi-conical structure representative of the ionized winds. The former region represents the accretion flow between the torus and the 
BLR, a necessary structure used to reproduce the observed polarization dichotomy in AGN \citep{Young2000,Goosmann2007}. The equatorial region is 
a flared disk with half-opening angle 20$^\circ$ from the equatorial plane and models a radiation-supported disk structure. It spans from 0.03 to 0.05~pc 
and is filled with electrons, such that the V-band optical depth along the equator is 1 \citep{Marin2012}. This equatorial inflow should not be confused
with the BLR, where most of the material is neutral (as evidenced by fast variations in the neutral absorbing column seen in X-ray observations, e.g. 
\citet{Maiolino2010} or \citet{Netzer2013} for models). The second structure is an electron-filled, hourglass-shaped wind that is outflowing from
0.1 to 30~pc from the center of the model. The optical depth of the wind is fixed to 0.05, a representative value of the optically-thin polar regions 
observed in nearby Seyferts \citep{Miller1991}. Increasing the optical depth contributed by the narrow-line region to values higher than 0.05
(such as presented in Fig.~3 in \citet{Schnorr2016}, where a typical A$_{\rm V}$ of 1 -- 2 mag is found) will not change the outcome of the 
simulation as long as the optical depth remains lower than unity. Otherwise multiple scattering in a Thomson-thick electron medium will increase the 
number of scatterings needed by radiation to escape the model along Seyfert-1 viewing angles, rotating the final polarization position angle by 90$^\circ$ 
(see \citet{Marin2012} for a grid of AGN polarized simulations with the polar winds V-band optical depth ranging from 0.01 to 3).

Our grid of results are presented in Fig.~\ref{Fig:Table_AGN_PO} and Fig.~\ref{Fig:Table_AGN_PA} for the polarization degree and the polarization 
position angle, respectively. Focusing on Seyfert-1 inclinations, we see that both $P$ and the $PPA$ are azimuthally-independent. The polarization degree 
resulting from scattering in a complex AGN model with a warped torus is of the order of 0.4\%. Scattering off the radiation-supported, inner disk provides most 
of the polarized information and the amount of secondary or tertiary reprocessing onto the dusty equatorial structure is not sufficient to impact the net 
$P$, nor the $PPA$, which is fixed to 90$^\circ$ (parallel polarization). Electrons in the polar winds do not contribute to $P$ as the medium is optically-thin
and forward scattering leads to small degrees of polarization. This type of signature is the same that was found by previous authors when modeling smooth-density 
or clumpy distributions of matter in complex AGN models \citep{Kartje1995,Marin2012,Marin2015}. This is also the typical amount of polarization found in nearby, 
equatorial-scattering dominated Seyferts \citep{Smith2002}. 

The case of an intermediate inclination is similar: the polarization position angle is almost always parallel to the symmetry axis of the model, with 
polarization degrees between 2 and 3\%. However, in rare cases, the polarization angle rotates by $\sim$~90$^\circ$, driving oscillations in the polarization 
degree (see, e.g., the case R$_{\rm warp}$ = 3~pc and $\theta_{\rm warp}$ = 30$^\circ$ in Fig.~\ref{Fig:Table_AGN_PO} and \ref{Fig:Table_AGN_PA}). This 
phenomenon is due to the occultation of the central region by a single or a group of dusty clouds. Scattered radiation from the wind will then prevail and 
rotate the $PPA$, impacting $P$. The polarization degree first decreases when the $PPA$ rotation occurs, then reaches a maximum ($<$~10\%) then decreases 
again when the $PPA$ stabilizes again. Most of those $PPA$ variations occur at an azimuthal angle close to 0$^\circ$, when the upper part of the warp (with
positive $z$ coordinates) can intercept the observer's viewing angle.

The case seems more complex for Seyfert-2 inclinations. If the polarization degree is indeed high due to Thomson scattering in the polar winds, leading to 
$PPA$ = 0$^\circ$ such as expected from Seyfert-2s \citep{Antonucci1993}, it may also show orthogonal rotations when the clump distribution is not dense 
enough to entirely cover the central region. Scattered radiation from the equatorial plane will dominate and rotate the polarization angle, decreasing the 
net polarization degree (the inverse phenomenon described for intermediate inclinations). However, variations in $P$ are not always related to $PPA$ 
oscillations: the warped structure, extending higher from the equatorial plane than a regular torus, may contribute to the global polarization by scattering 
photons that emerge from the dust funnel or from the radiation-supported disk. The mainly parallel polarization angle of those photons will mix with 
the perpendicular signature of photon scattered in the polar region and thus cancel $P$. The contribution of the radiation with parallel $PPA$ is weaker, 
thus $P$ does not completely cancel out, but might be significant enough to bring the polarization degree to a couple of percents. Coupled to the clumpiness 
effect and the partial covering of the nucleus, $P$ can be as low as 0.2\%. This case is statistically very unlikely to happen, otherwise AGN with optical 
Seyfert-2 classifications but Seyfert-1 polarization degree and angle would have been observed yet \citep{Antonucci1993}.

\subsection{Imaging polarimetry}
\label{AGN:Maps}

\begin{figure*}[ht] 
   \setcounter{figure}{7}
   \begin{minipage}[b]{0.33\linewidth}
      \centering
      \includegraphics[trim = 0mm 4.1mm 5mm 0mm, clip, width=5.92cm]{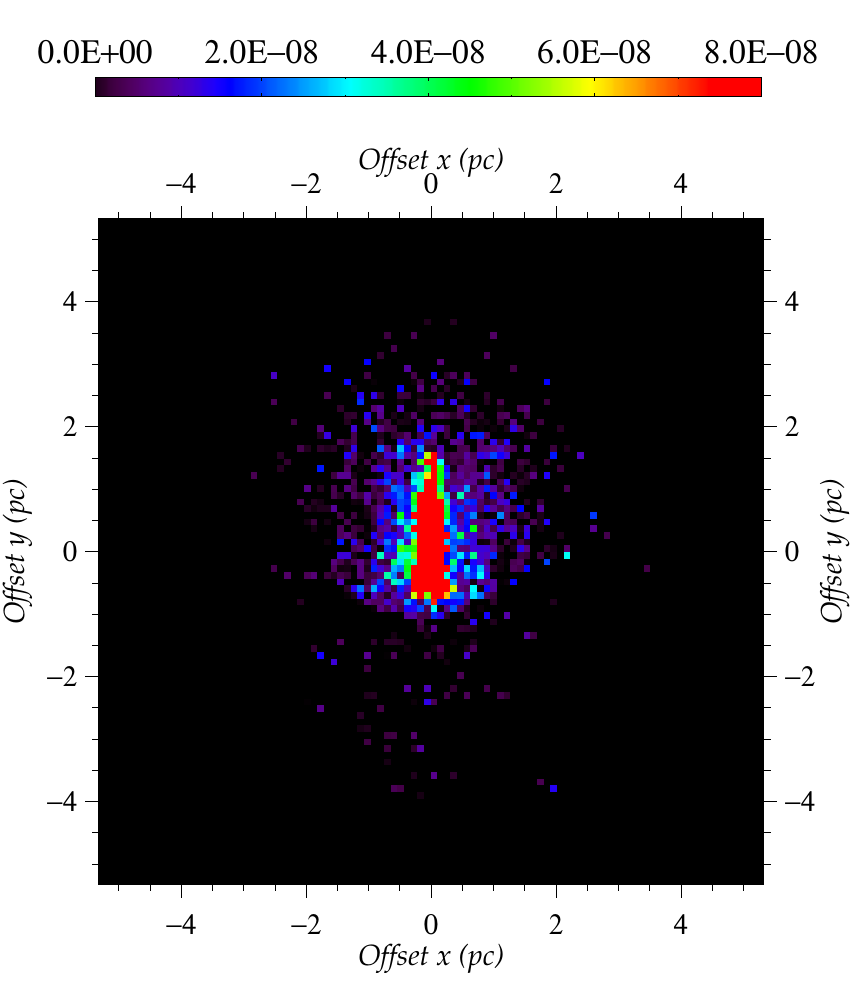} 
   \end{minipage} 
   \begin{minipage}[b]{0.33\linewidth}
      \centering   
      \includegraphics[trim = 4mm 4.1mm 5mm 0mm, clip, width=5.65cm]{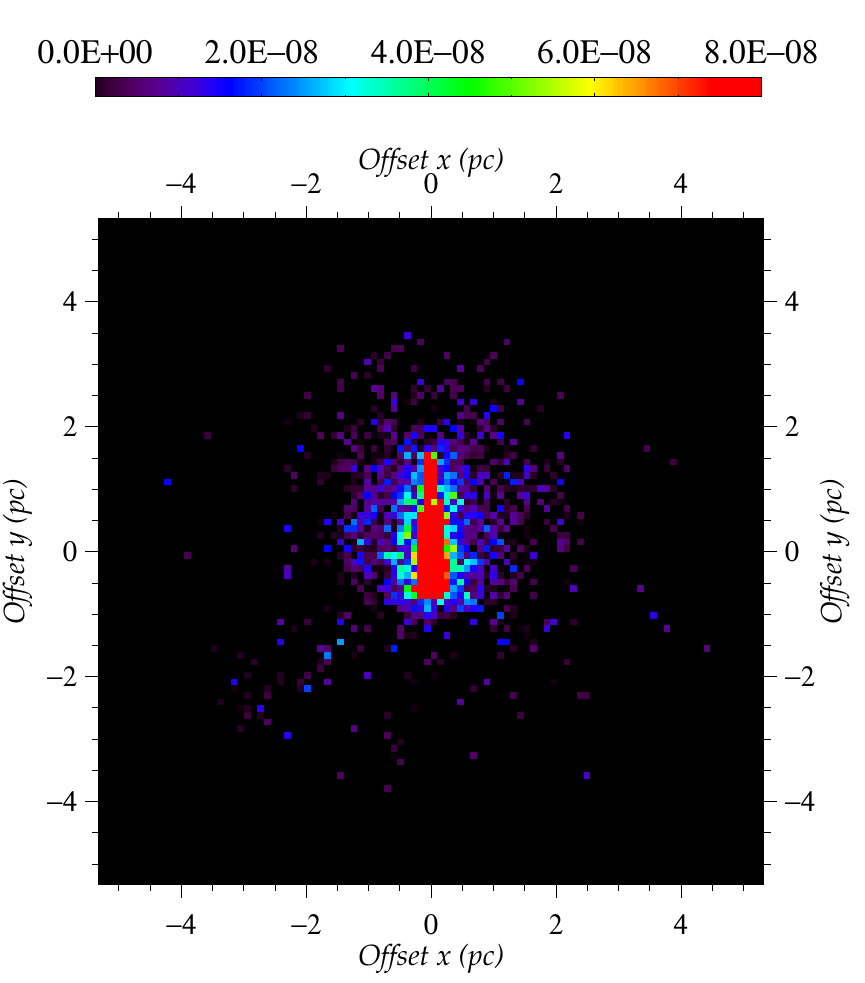} 
   \end{minipage}
   \begin{minipage}[b]{0.33\linewidth}
      \centering   
      \includegraphics[trim = 4mm 4.1mm 0mm 0mm, clip, width=6.01cm]{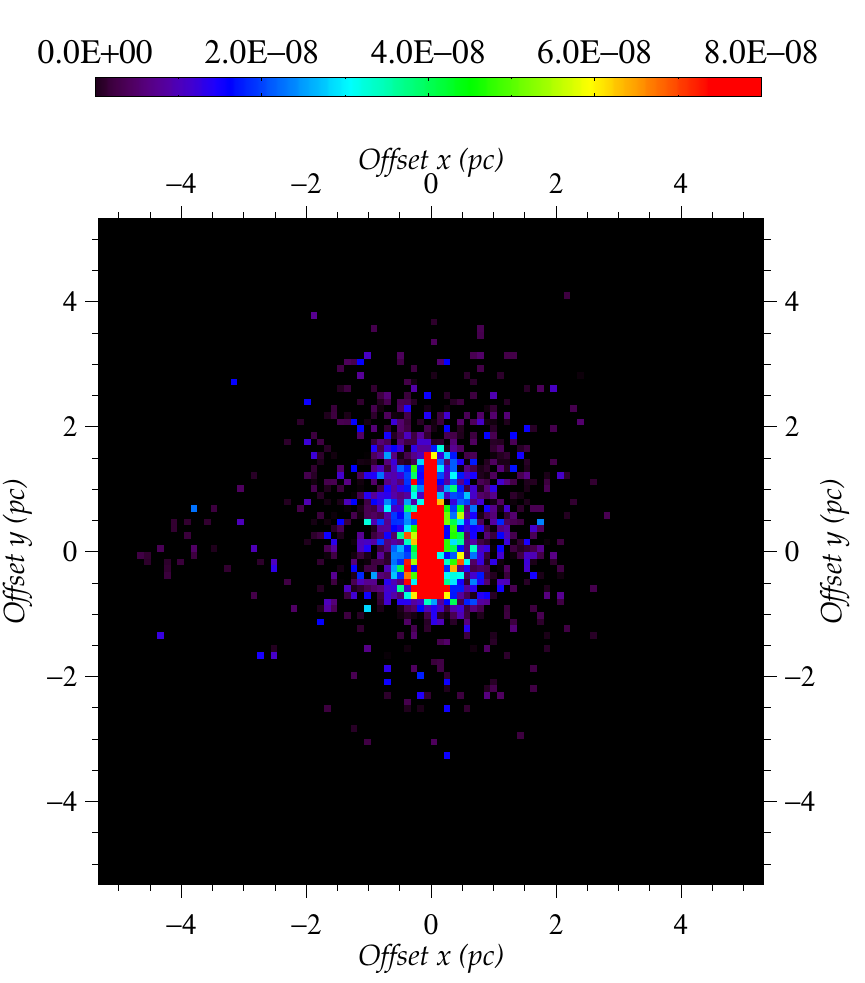} 
   \end{minipage}   
   \begin{minipage}[b]{0.33\linewidth}
      \centering
      \includegraphics[trim = 0mm 4.1mm 5mm 21.5mm, clip, width=5.92cm]{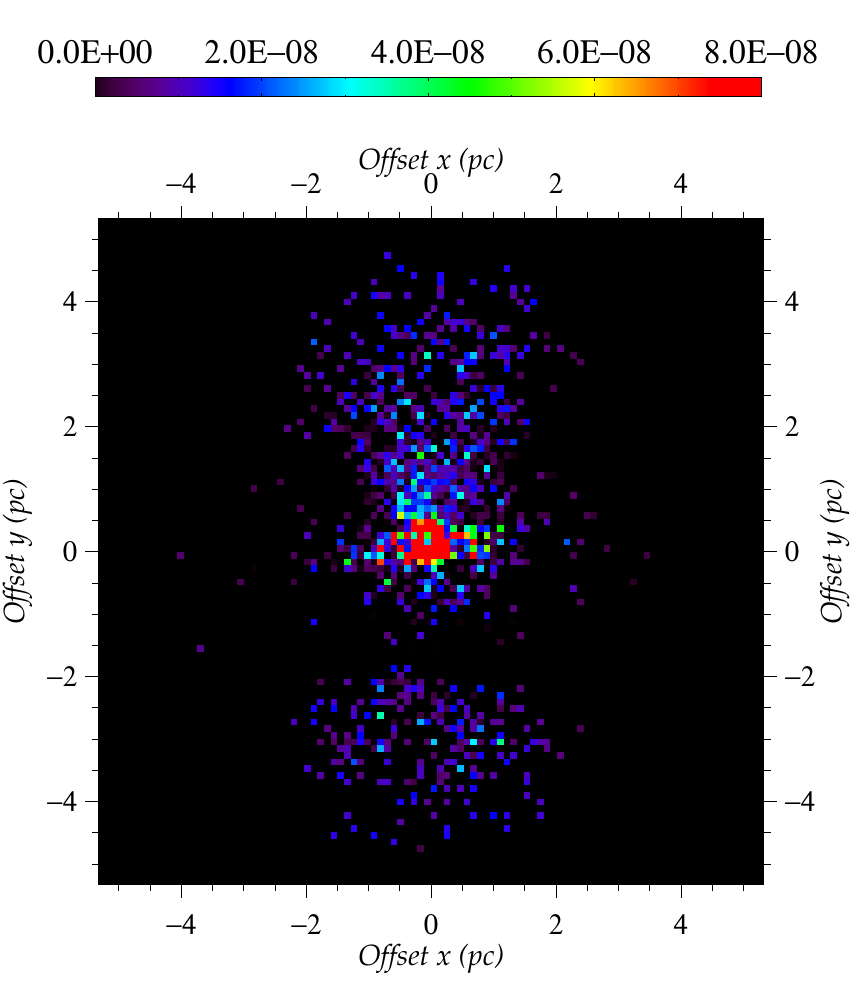} 
   \end{minipage} 
   \begin{minipage}[b]{0.33\linewidth}
      \centering   
      \includegraphics[trim = 4mm 4.1mm 5mm 21.5mm, clip, width=5.65cm]{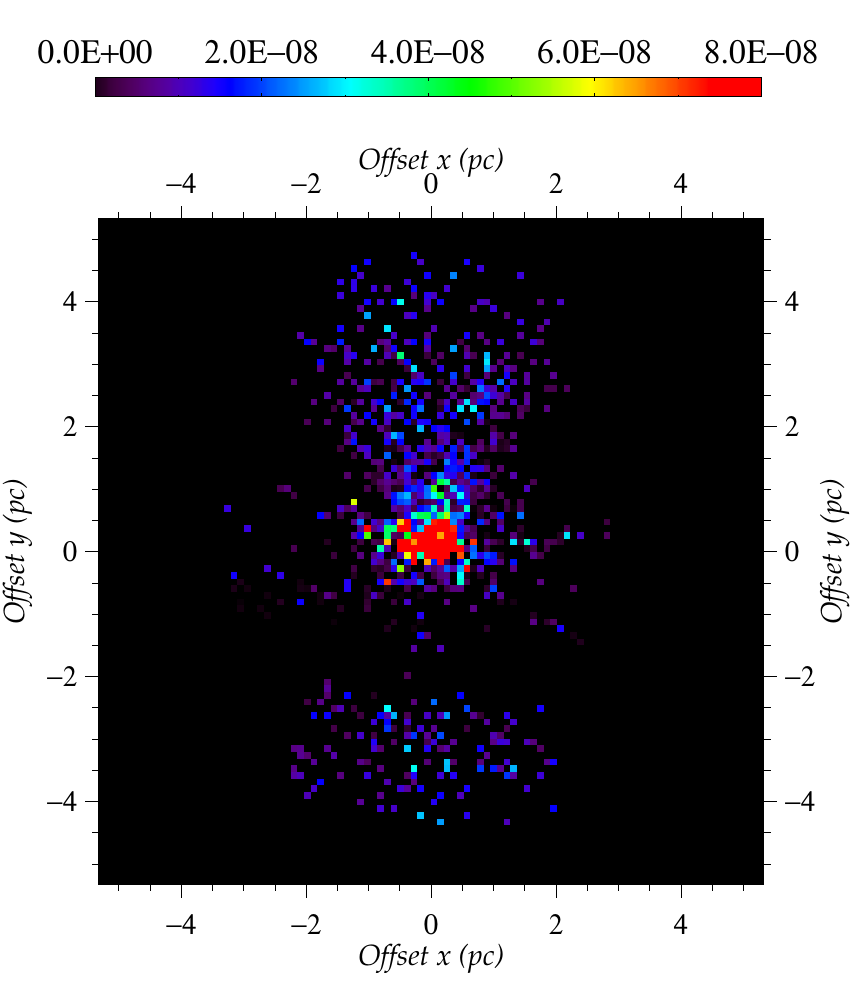}  
   \end{minipage}
   \begin{minipage}[b]{0.33\linewidth}
      \centering   
      \includegraphics[trim = 4mm 4.1mm 0mm 21.5mm, clip, width=6.01cm]{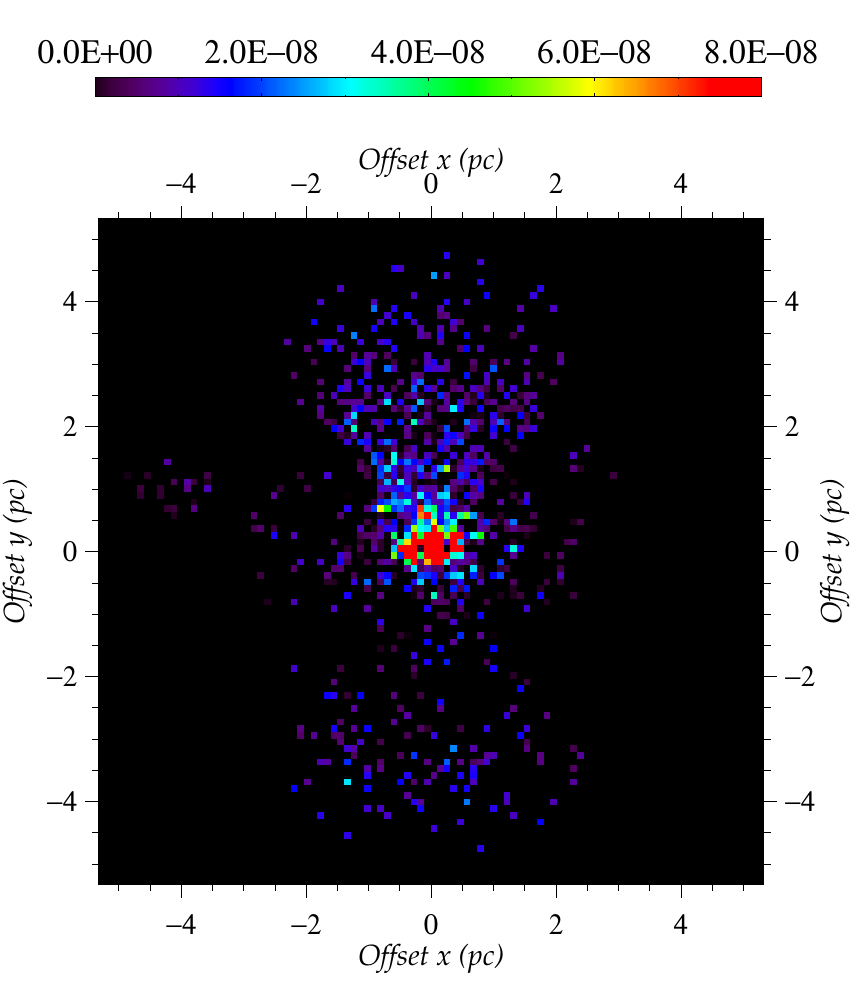} 
   \end{minipage} 
   \begin{minipage}[b]{0.33\linewidth}
      \centering
      \includegraphics[trim = 0mm 0mm 5mm 21.5mm, clip, width=5.99cm]{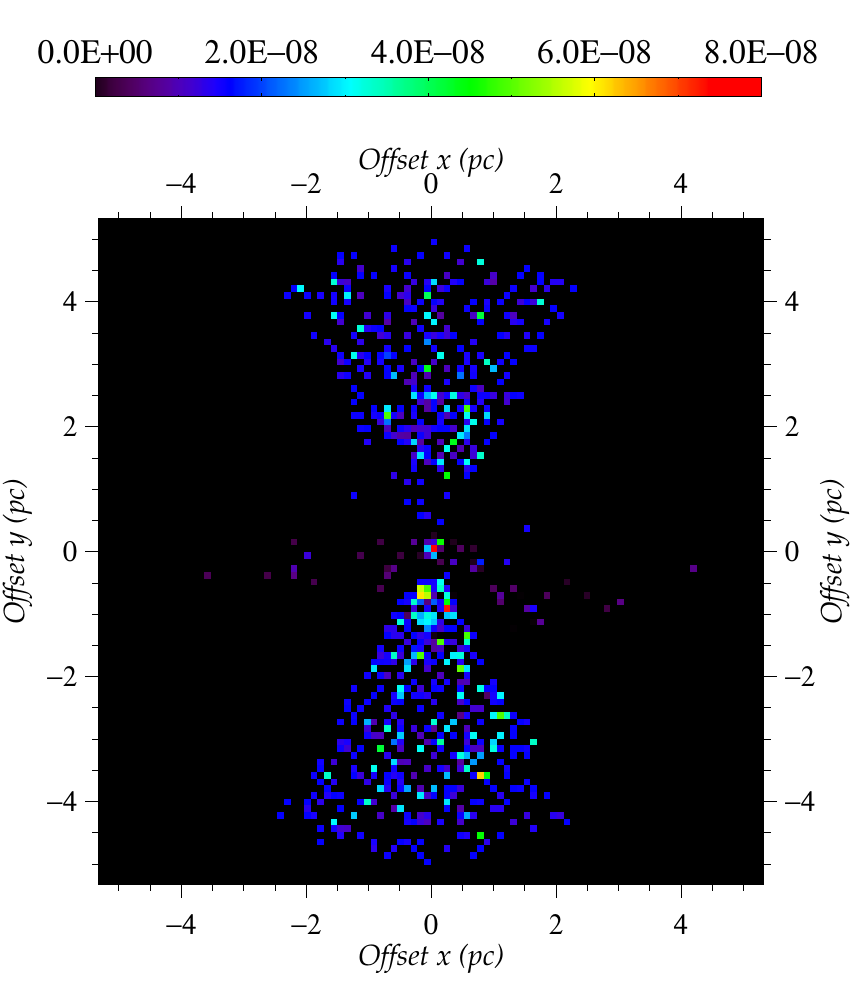} 
   \end{minipage} 
   \begin{minipage}[b]{0.33\linewidth}
      \centering   
      \includegraphics[trim = 5mm 0mm 5mm 21.5mm, clip, width=5.65cm]{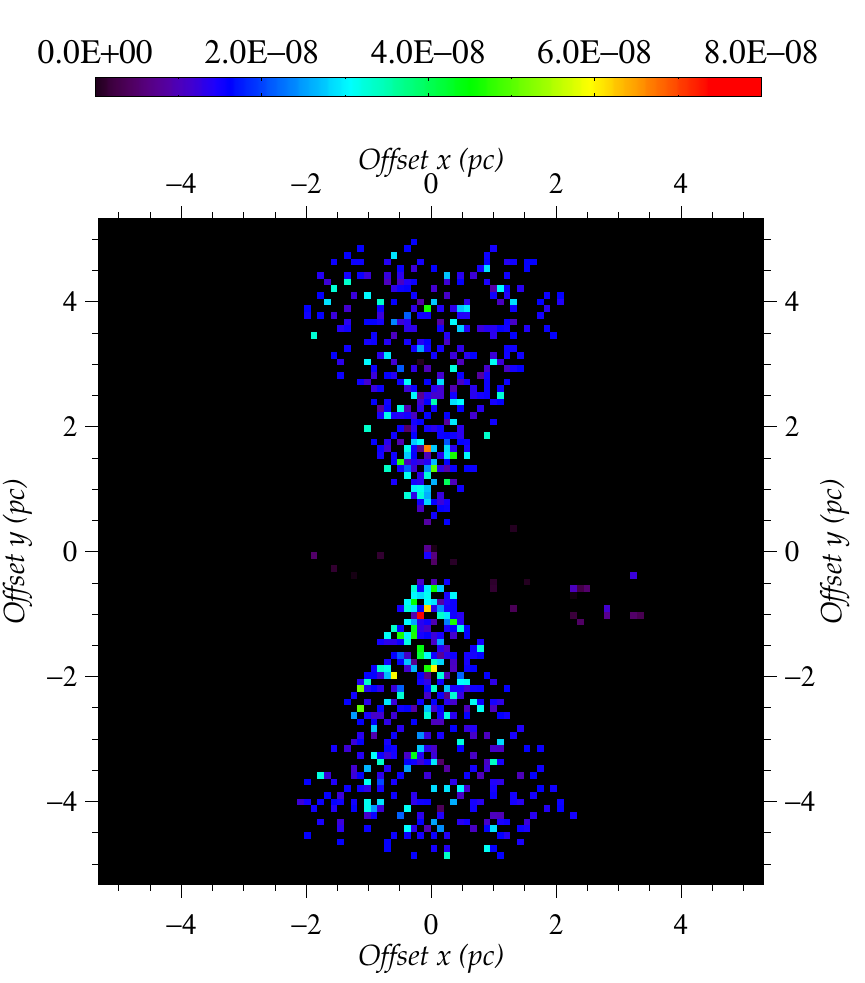} 
   \end{minipage}
   \begin{minipage}[b]{0.33\linewidth}
      \centering   
      \includegraphics[trim = 5mm 0mm 0mm 21.5mm, clip, width=6.01cm]{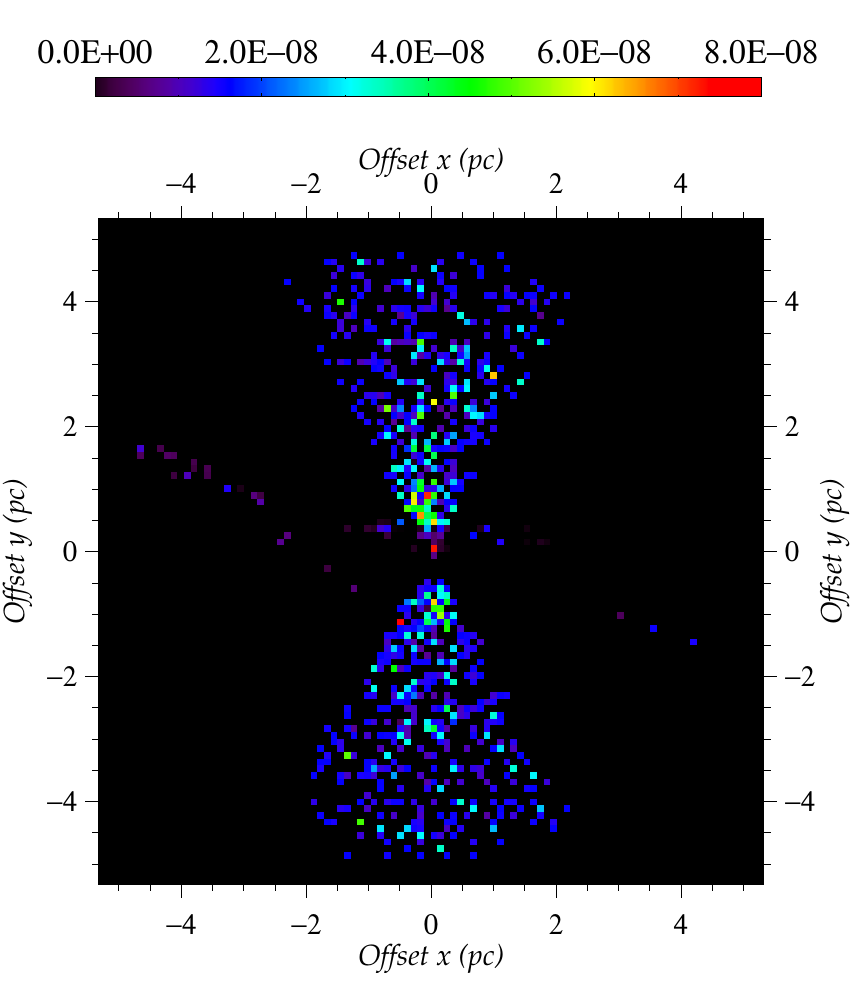} 
   \end{minipage}    
   \caption{Polarized flux maps of a complex AGN model using a warped and 
	    clumpy dusty torus with R$_{\rm warp}$ = 2~pc and 
	    $\theta_{\rm warp}$ = 30$^\circ$, such as in Fig.~\ref{Fig:3Dplots}.
	    The polarized flux is color coded (in arbitrary units, but 
	    identically scaled for all panels). The axes are in parsecs. 
	    The first row shows a nucleus orientation of 18$^\circ$, 
	    the second row an inclination of 50$^\circ$ and the last 
	    row 87$^\circ$. Each column shows a different azimuthal 
	    angle: 0$^\circ$, 45$^\circ$, and 90$^\circ$.}
   \label{Fig:Maps_TF}
\end{figure*}

\begin{figure*}[ht] 
  \setcounter{figure}{8}
   \begin{minipage}[b]{0.33\linewidth}
      \centering
      \includegraphics[trim = 0mm 4.1mm 5mm 0mm, clip, width=5.92cm]{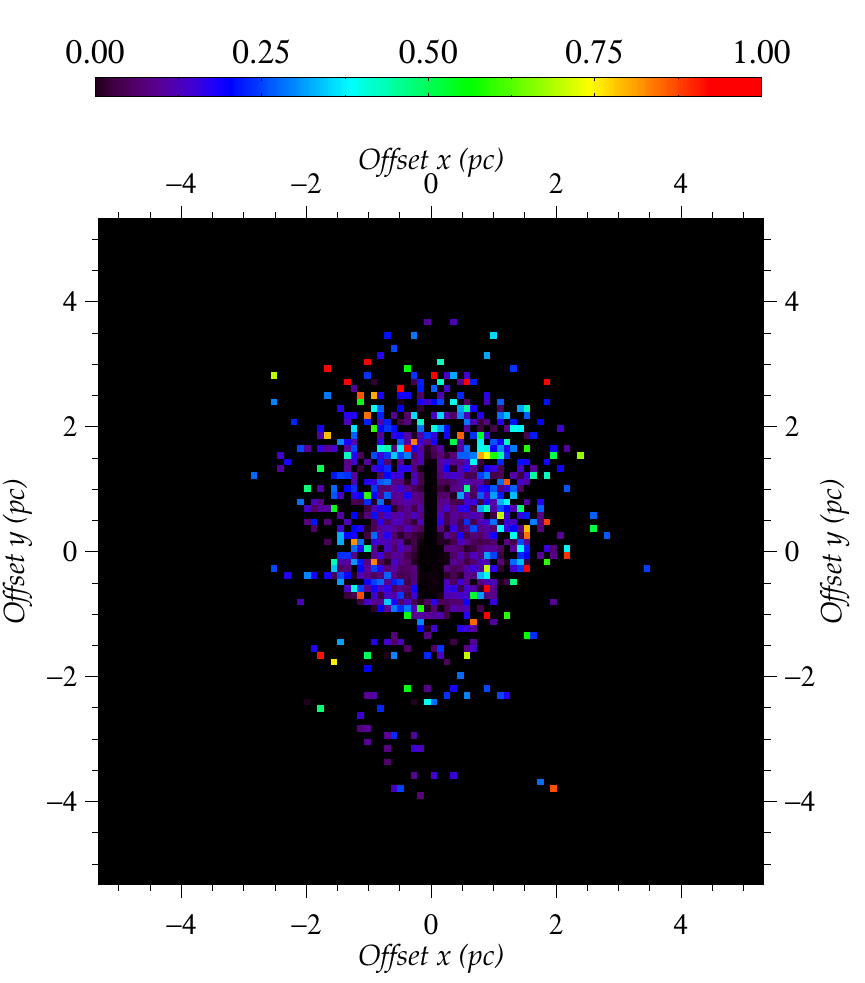} 
   \end{minipage} 
   \begin{minipage}[b]{0.33\linewidth}
      \centering   
      \includegraphics[trim = 4mm 4.1mm 5mm 0mm, clip, width=5.65cm]{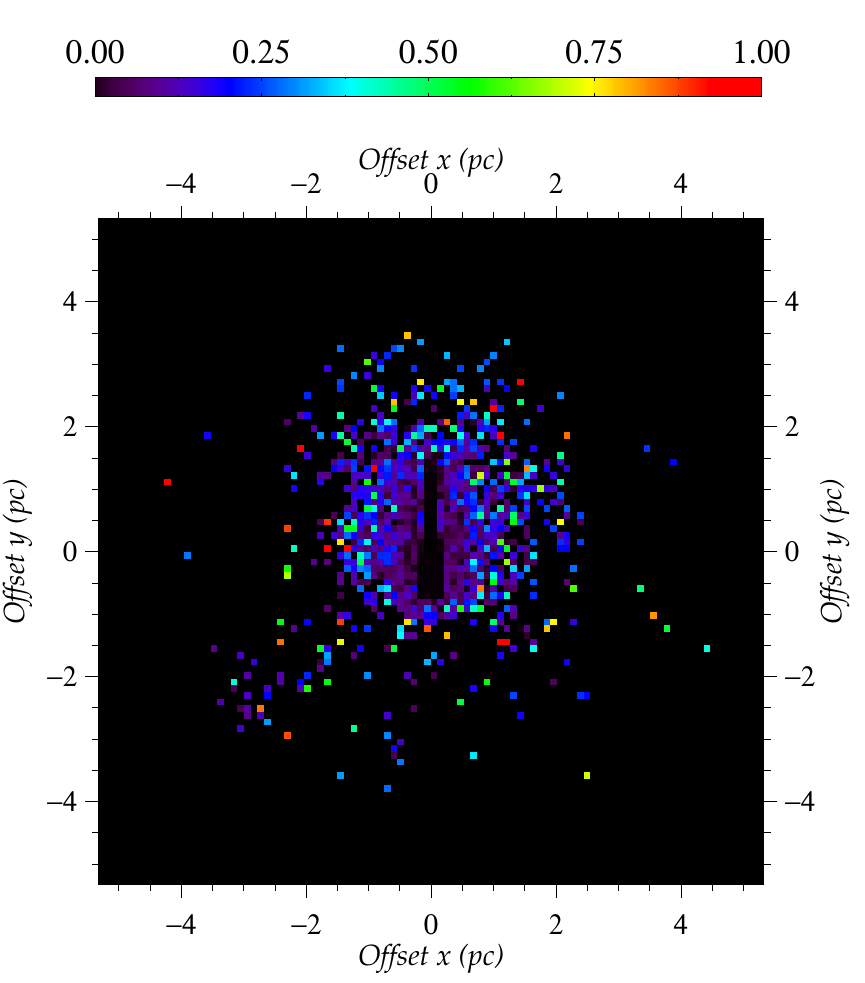} 
   \end{minipage}
   \begin{minipage}[b]{0.33\linewidth}
      \centering   
      \includegraphics[trim = 4mm 4.1mm 0mm 0mm, clip, width=6.01cm]{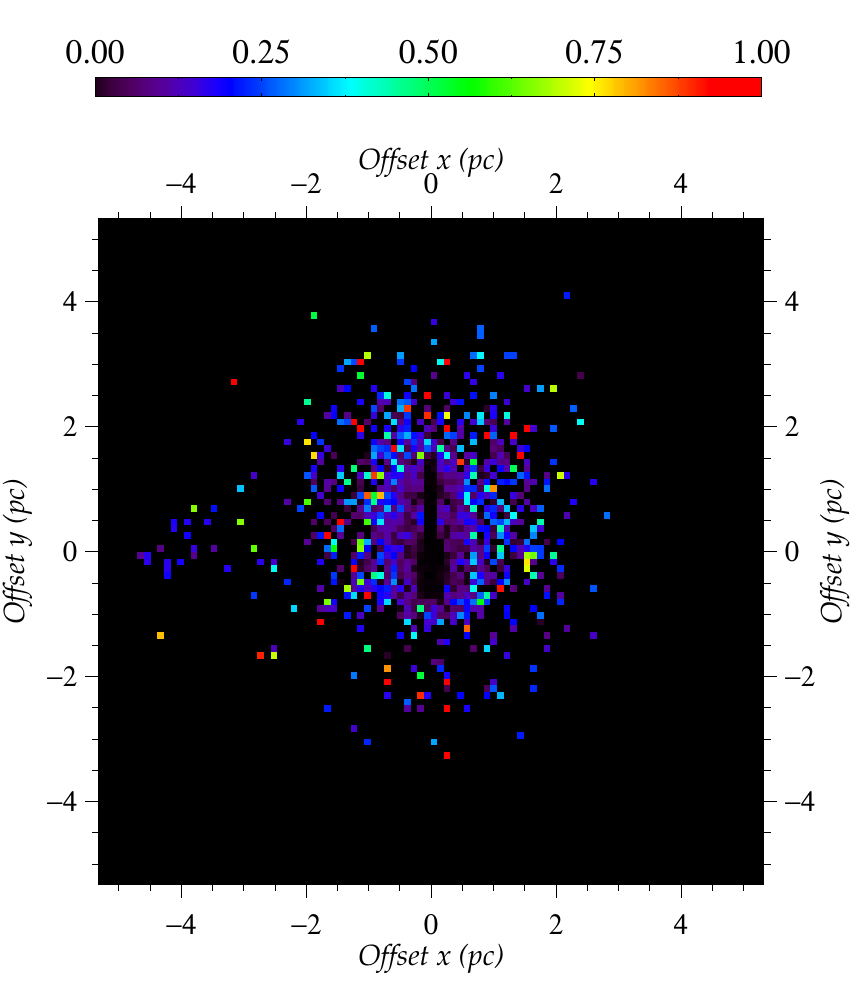} 
   \end{minipage}   
   \begin{minipage}[b]{0.33\linewidth}
      \centering
      \includegraphics[trim = 0mm 4.1mm 5mm 21.5mm, clip, width=5.92cm]{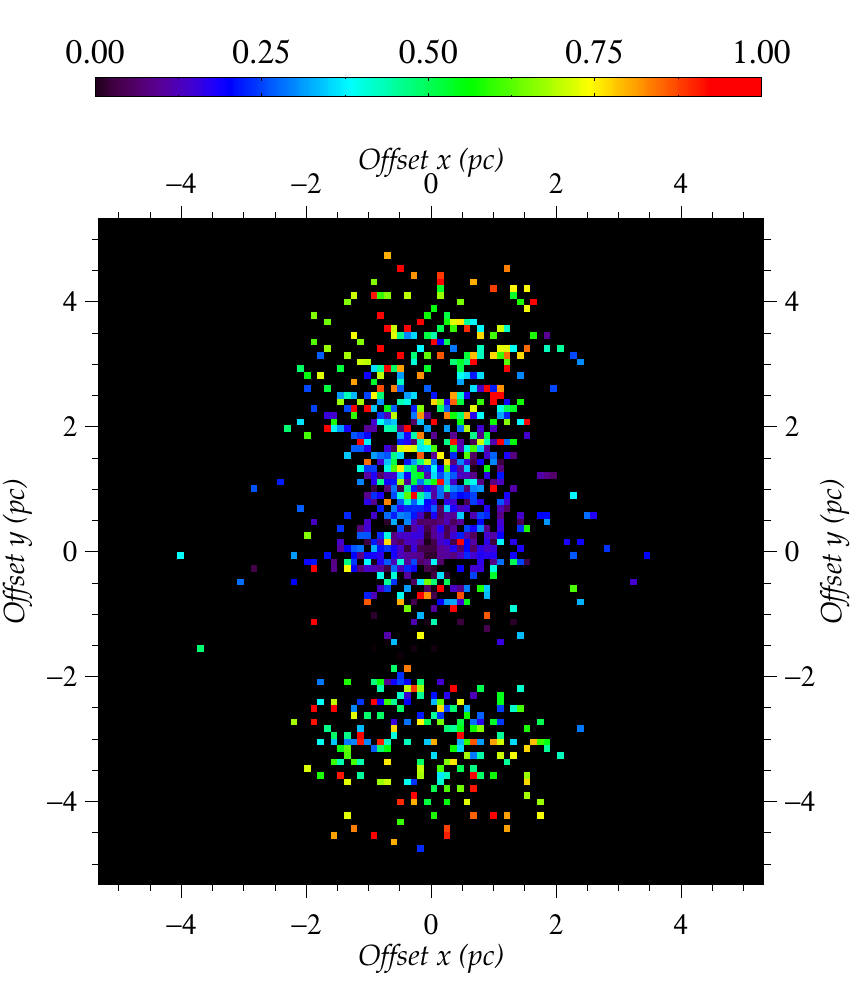} 
   \end{minipage} 
   \begin{minipage}[b]{0.33\linewidth}
      \centering   
      \includegraphics[trim = 4mm 4.1mm 5mm 21.5mm, clip, width=5.65cm]{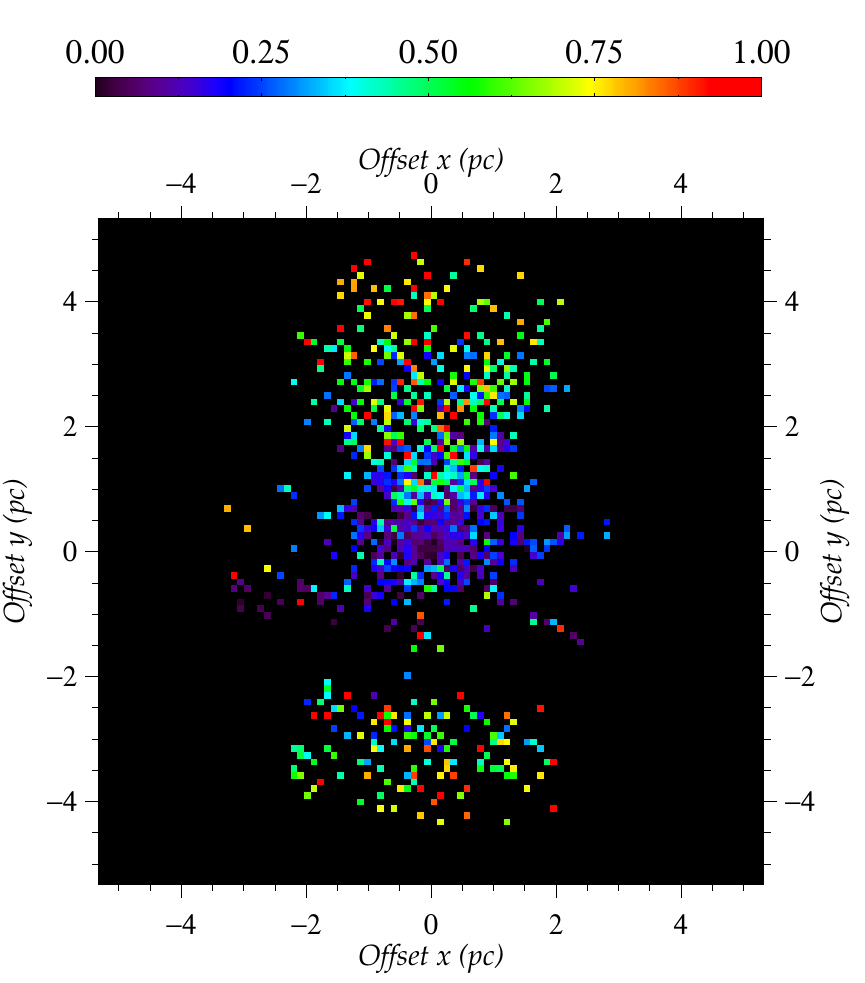}  
   \end{minipage}
   \begin{minipage}[b]{0.33\linewidth}
      \centering   
      \includegraphics[trim = 4mm 4.1mm 0mm 21.5mm, clip, width=6.01cm]{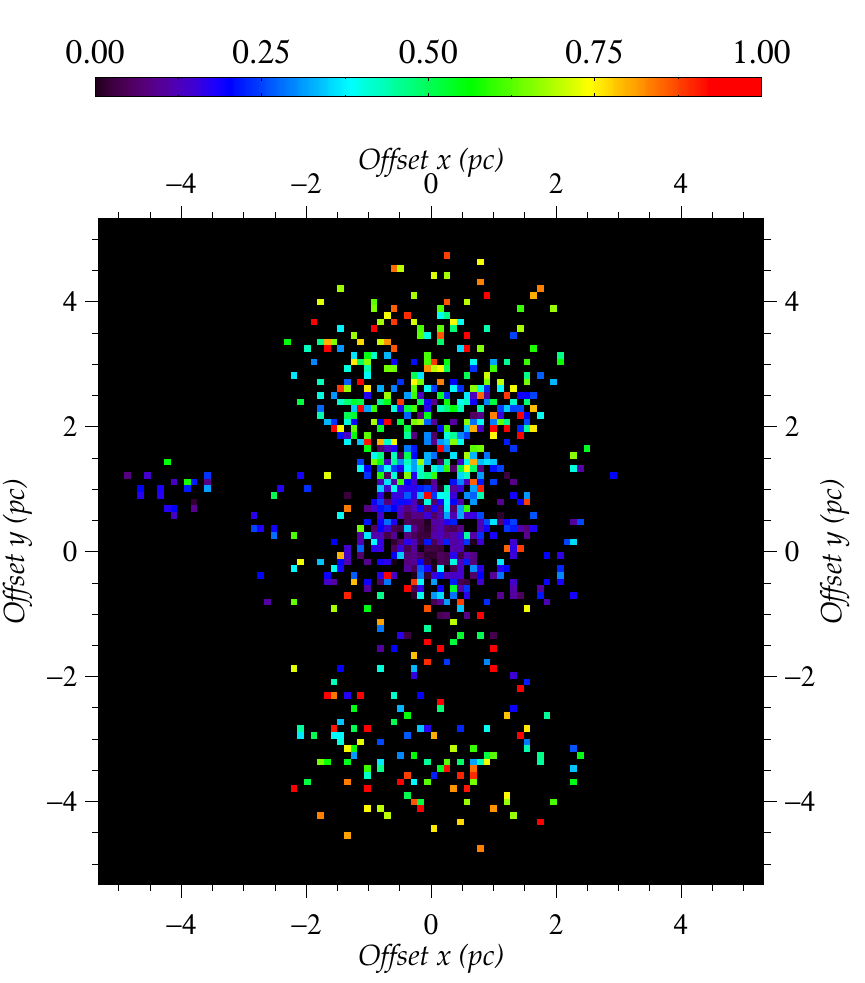} 
   \end{minipage} 
   \begin{minipage}[b]{0.33\linewidth}
      \centering
      \includegraphics[trim = 0mm 0mm 5mm 21.5mm, clip, width=5.99cm]{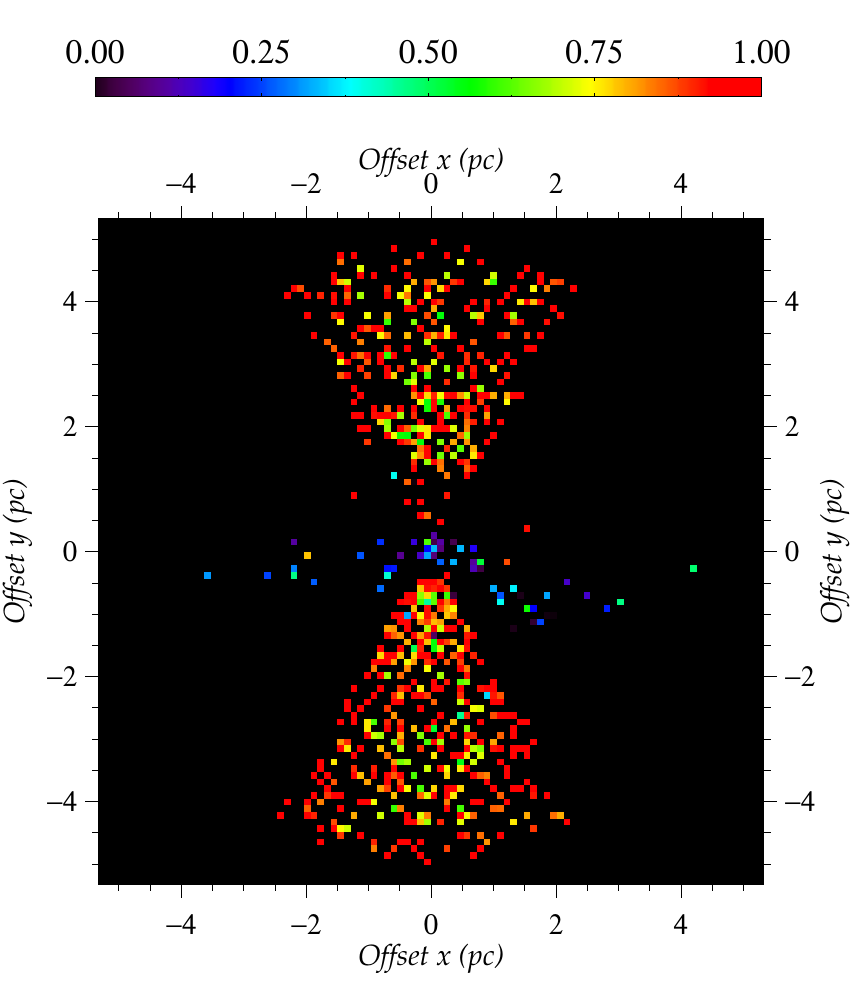} 
   \end{minipage} 
   \begin{minipage}[b]{0.33\linewidth}
      \centering   
      \includegraphics[trim = 5mm 0mm 5mm 21.5mm, clip, width=5.65cm]{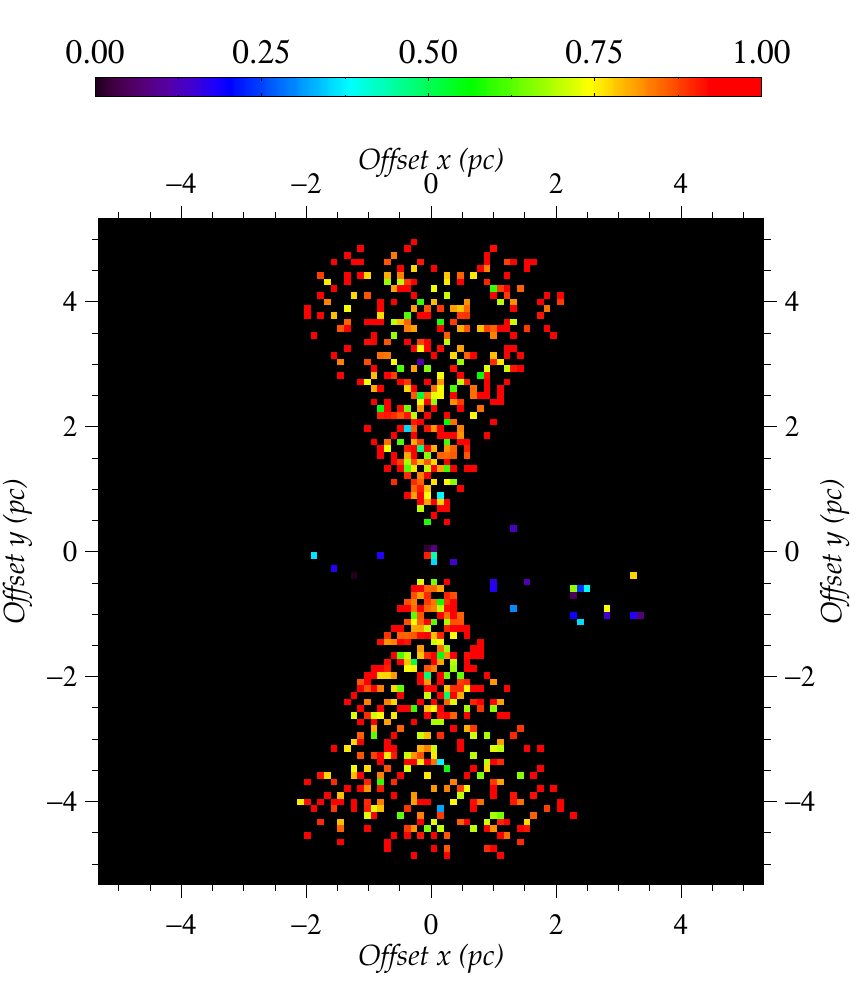} 
   \end{minipage}
   \begin{minipage}[b]{0.33\linewidth}
      \centering   
      \includegraphics[trim = 5mm 0mm 0mm 21.5mm, clip, width=6.01cm]{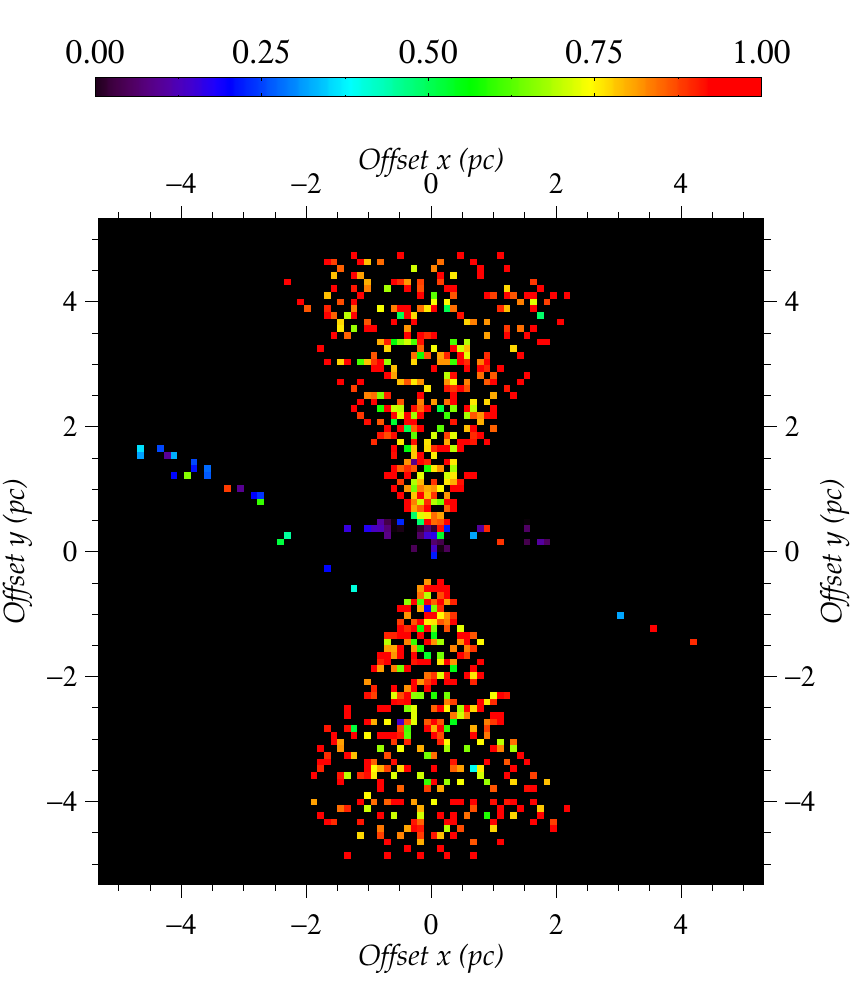} 
   \end{minipage}     
  \caption{Same as Fig.~\ref{Fig:Maps_TF} but showing the optical 
	   polarization fraction ranging from 0 (unpolarized) to 
	   1 (fully polarized).}
  \label{Fig:Maps_PO}
\end{figure*}

\begin{figure*}[ht] 
  \setcounter{figure}{9}
   \begin{minipage}[b]{0.33\linewidth}
      \centering
      \includegraphics[trim = 0mm 4.1mm 5mm 0mm, clip, width=5.92cm]{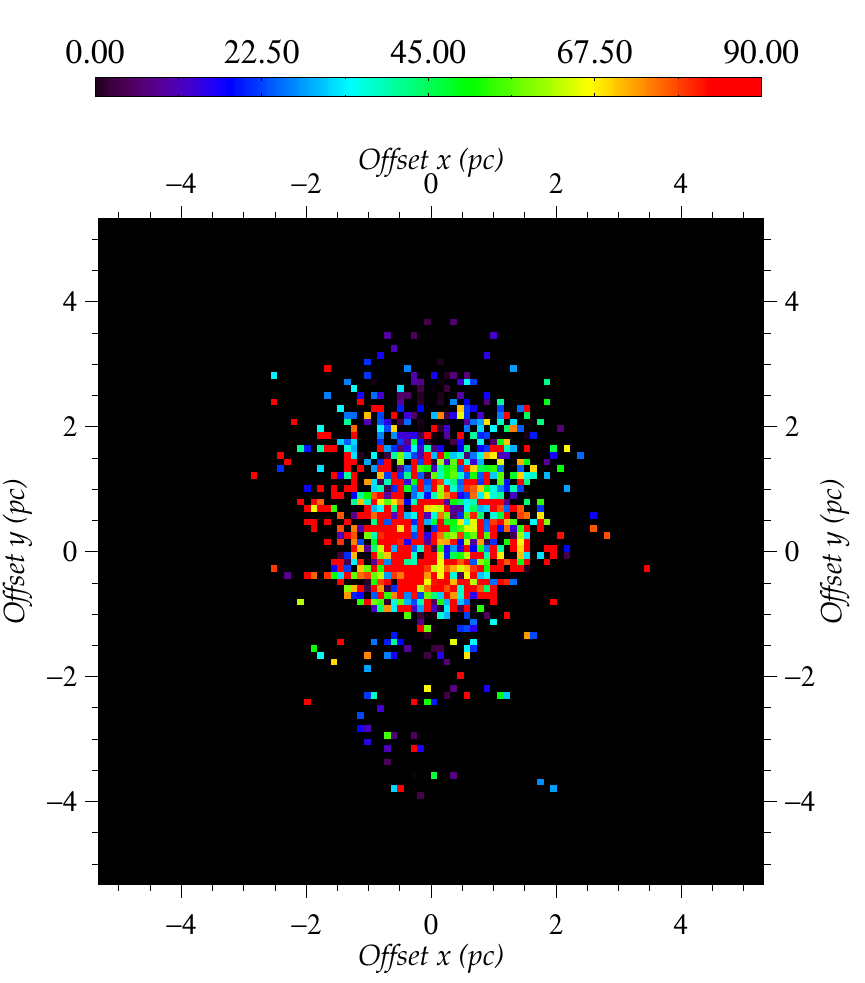} 
   \end{minipage} 
   \begin{minipage}[b]{0.33\linewidth}
      \centering   
      \includegraphics[trim = 4mm 4.1mm 5mm 0mm, clip, width=5.65cm]{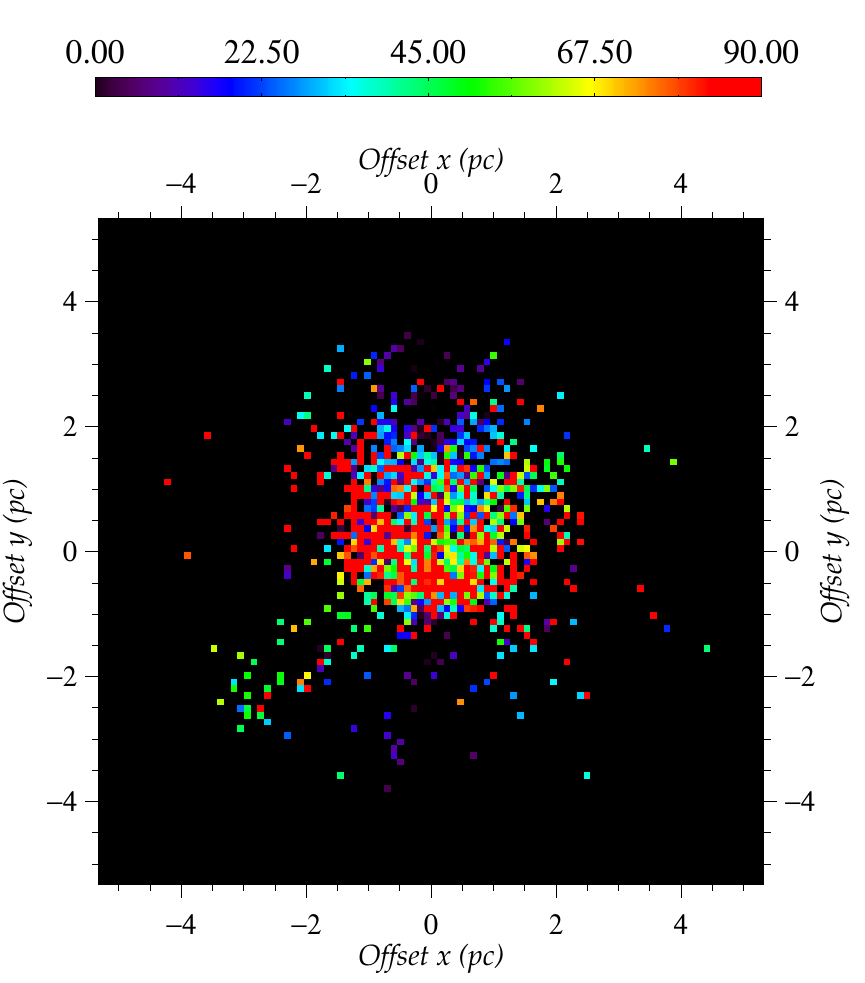} 
   \end{minipage}
   \begin{minipage}[b]{0.33\linewidth}
      \centering   
      \includegraphics[trim = 4mm 4.1mm 0mm 0mm, clip, width=6.01cm]{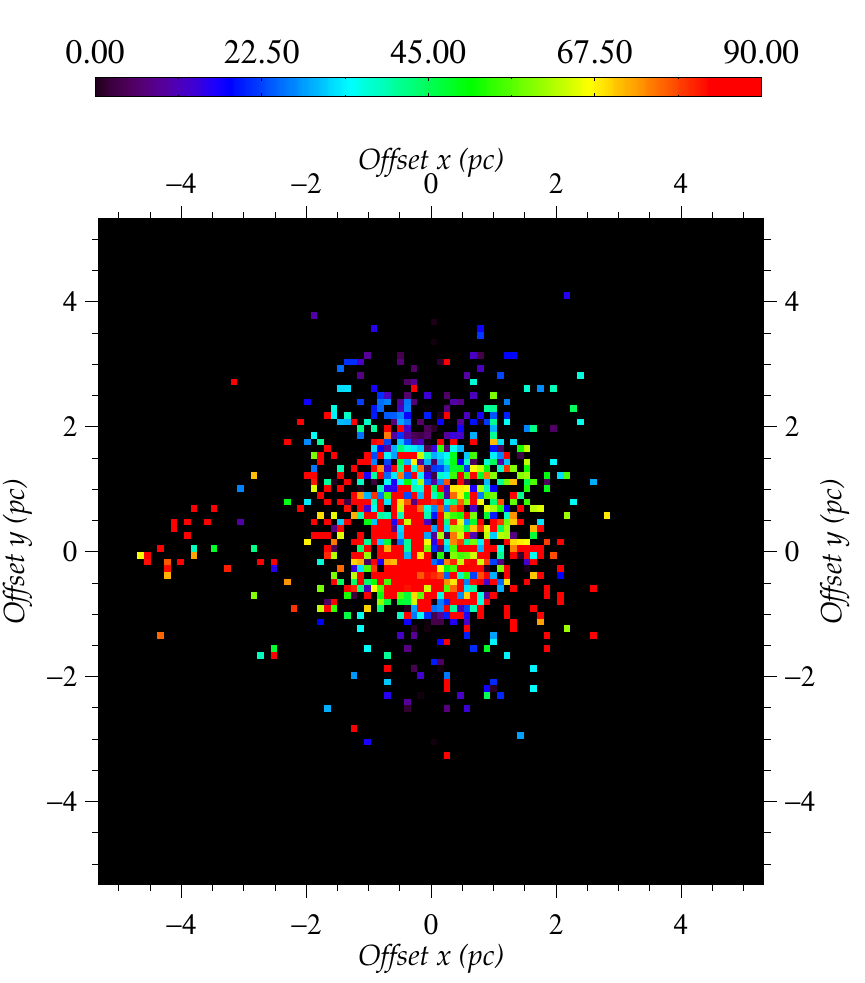} 
   \end{minipage}   
   \begin{minipage}[b]{0.33\linewidth}
      \centering
      \includegraphics[trim = 0mm 4.1mm 5mm 21.5mm, clip, width=5.92cm]{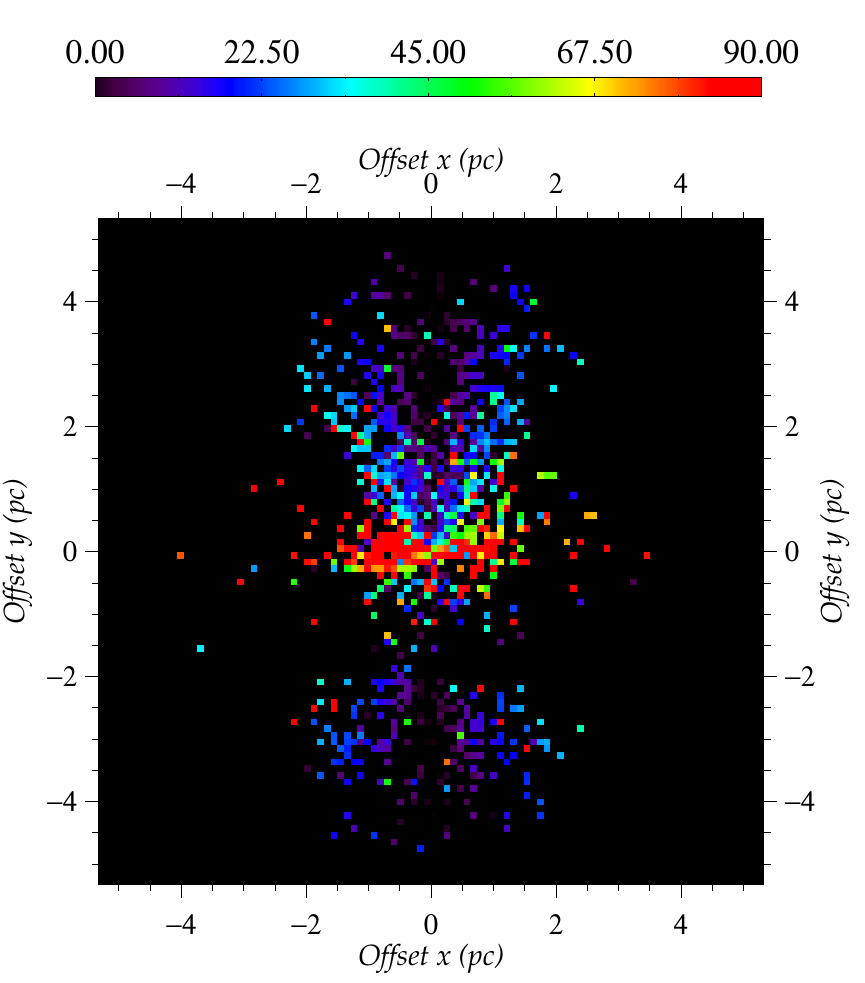} 
   \end{minipage} 
   \begin{minipage}[b]{0.33\linewidth}
      \centering   
      \includegraphics[trim = 4mm 4.1mm 5mm 21.5mm, clip, width=5.65cm]{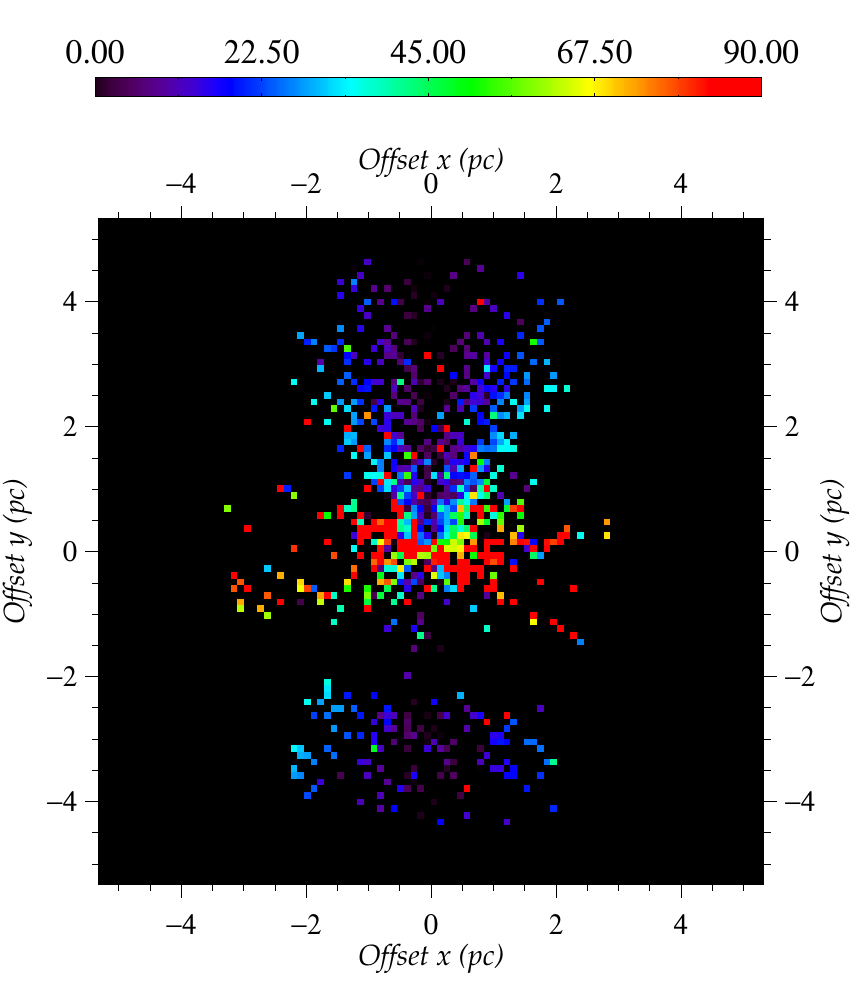}  
   \end{minipage}
   \begin{minipage}[b]{0.33\linewidth}
      \centering   
      \includegraphics[trim = 4mm 4.1mm 0mm 21.5mm, clip, width=6.01cm]{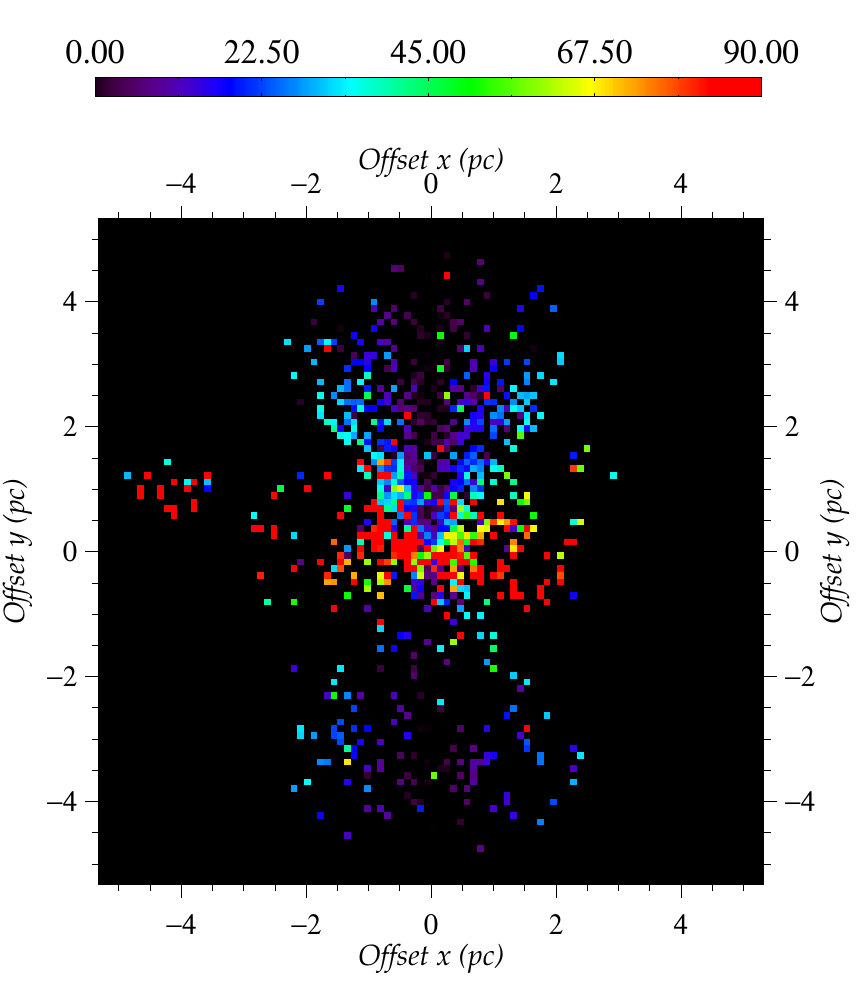} 
   \end{minipage} 
   \begin{minipage}[b]{0.33\linewidth}
      \centering
      \includegraphics[trim = 0mm 0mm 5mm 21.5mm, clip, width=5.99cm]{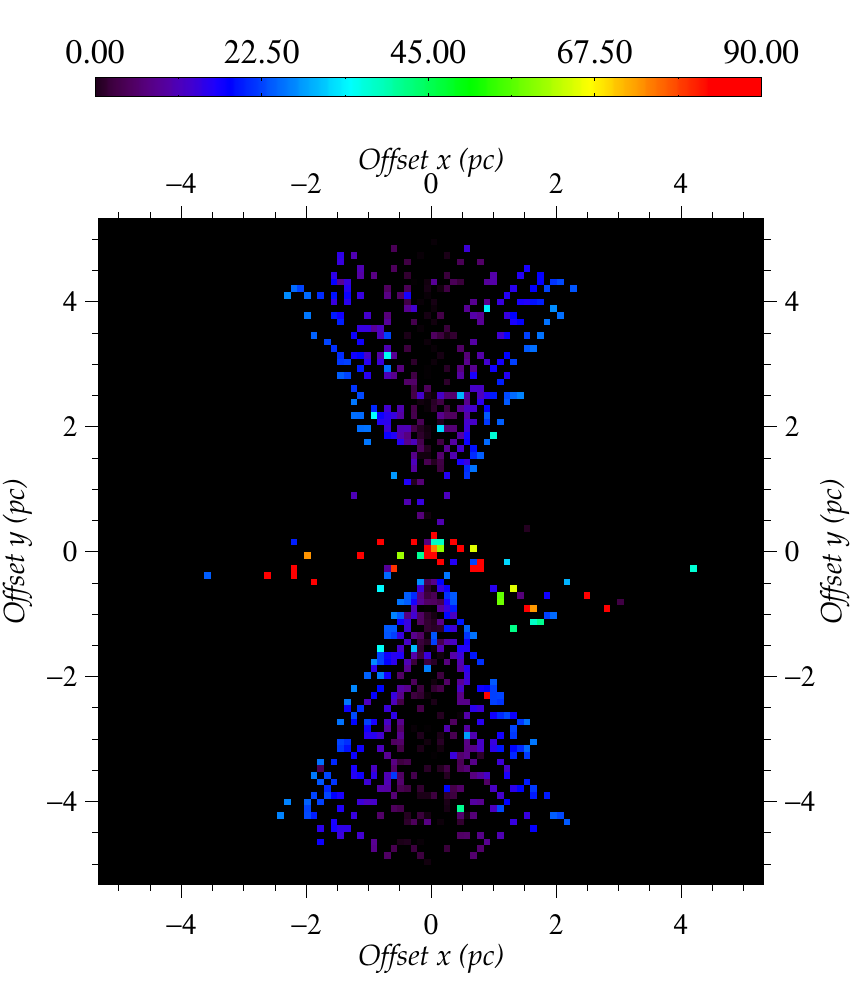} 
   \end{minipage} 
   \begin{minipage}[b]{0.33\linewidth}
      \centering   
      \includegraphics[trim = 5mm 0mm 5mm 21.5mm, clip, width=5.65cm]{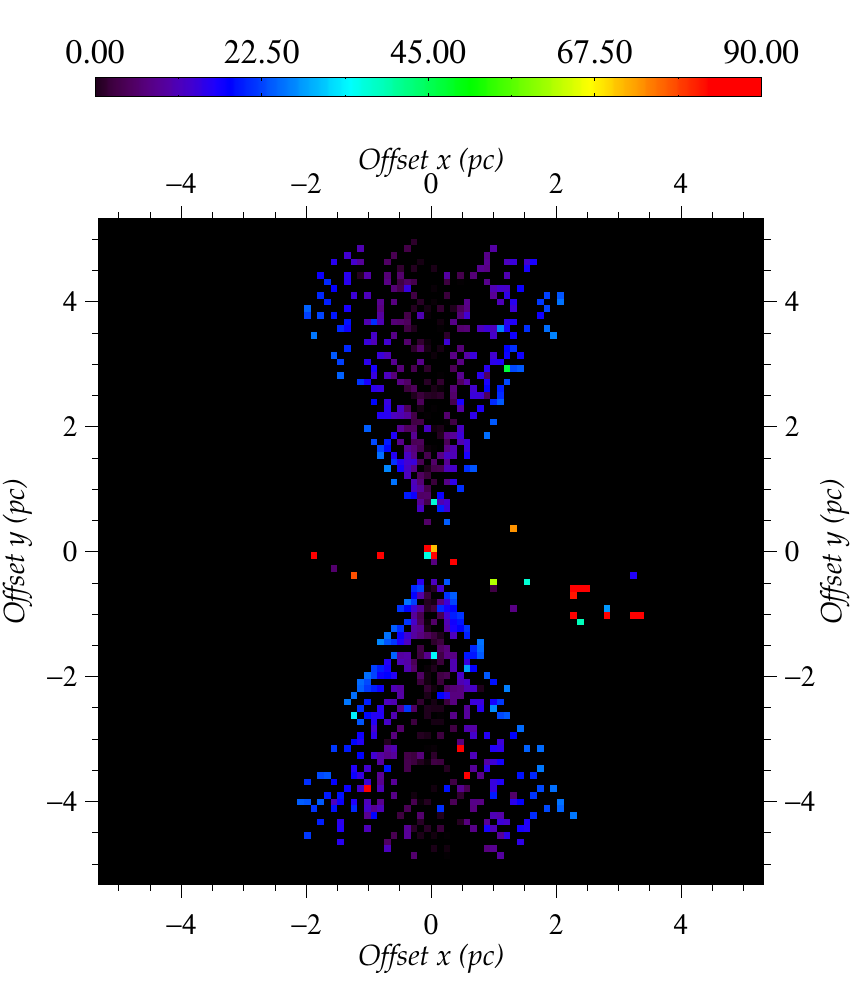} 
   \end{minipage}
   \begin{minipage}[b]{0.33\linewidth}
      \centering   
      \includegraphics[trim = 5mm 0mm 0mm 21.5mm, clip, width=6.01cm]{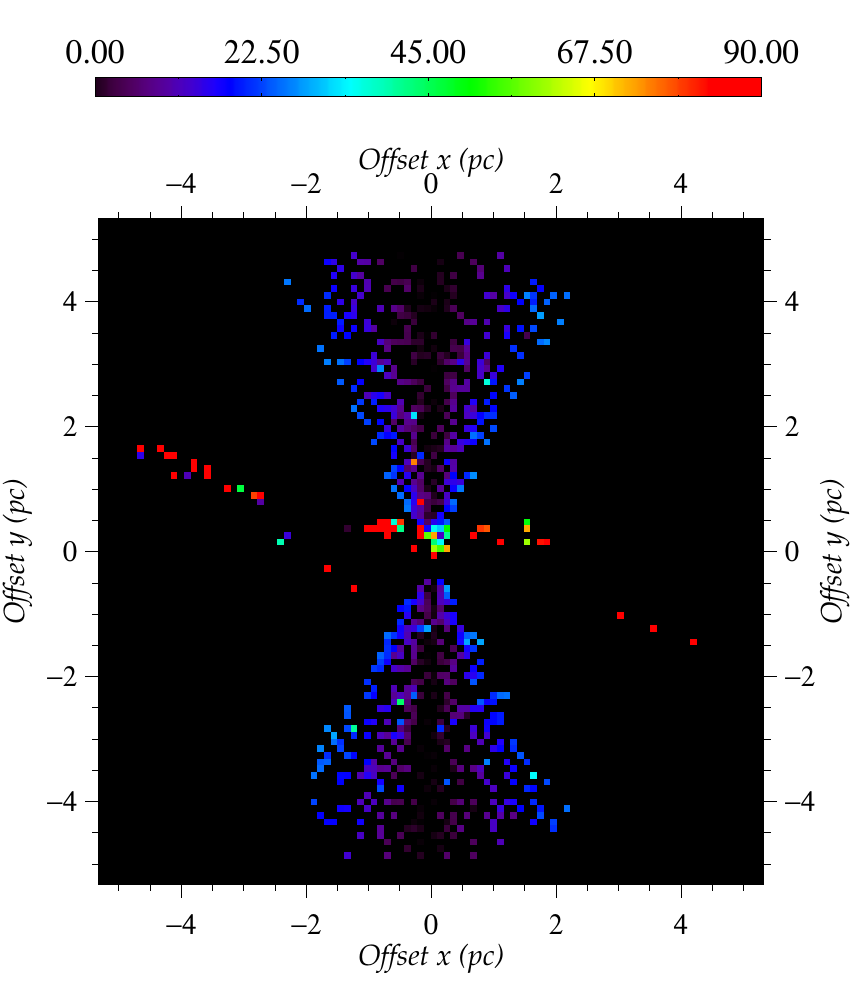} 
   \end{minipage}     
  \caption{Same as Fig.~\ref{Fig:Maps_TF} but showing the optical 
	   polarization position angle ranging from 0$^\circ$ (perpendicular polarization 
	   angle with respect to the projected radio axis of the system) to 
	   90$^\circ$ (parallel polarization angle).}
  \label{Fig:Maps_PA}
\end{figure*}

Revealing the polarization pattern around the hidden nuclei of Seyfert-2s was a major success when the polarimeter
on-board of the \textit{Hubble Space Telescope} was available and helped the community to better understand the morphology 
and composition of the resolvable structures surrounding the obscured SMBH. In particular, imaging polarimetry helped to 
localize the position of the hidden source of UV continuum, highlighted the shape of the ionized polar winds, detected 
dichroic signatures in linear polarization, estimated the amount of foreground extinction, examined the composition of the NLR 
and lead to refined AGN models by better constraining the distribution of the scattering medium (see \citealt{Capetti1995,Capetti1996},
and more recently \citealt{Gratadour2015} with the Spectro-Polarimetric High-contrast Exoplanet REsearch instrument 
(SPHERE) on the \textit{Very Large Telescope}, VLT). Polarization 
induced by dust scattering onto the circumnuclear region has been undetectable so far and will stay impossible to resolve 
for many years due to the small amount of flux reprocessed by this region and due to the technical limitations of current 
facilities. As an example, under decent atmospheric conditions, a maximum of 60~milliarcsecond ($\sim$~4~pc) resolution 
for imaging polarimetry is achievable on NGC~1068 ($d$ = 14.4~Mpc) in the H and K bands using adaptive optics \citep{Gratadour2015}. 
On the other hand, General Relativity Analysis via VLT InTerferometrY (GRAVITY) at the VLTI (in the K band) has an 
imaging resolution of about 2~mas and may be able to measure polarization in a few bright AGN thanks to its Wollaston prism 
\citep{Gillessen2006}. Yet, observations will be limited to very few sources and the polarized signal from the torus might 
be diluted by the emission from the cones. However, if the torus is warped and extended along the polar regions, it might 
be possible to detect some effects on the AGN polarization maps.

To investigate the impact of warped tori, we ran the AGN model presented in Sect.~\ref{AGN:Tables} using the following warp 
parametrization: R$_{\rm warp}$ = 2~pc and $\theta_{\rm warp}$ = 30$^\circ$. We created the polarization maps of the AGN at 
the same three viewing angles as shown in all the figures of this paper (18$^\circ$, 50$^\circ$ and 87$^\circ$), together with 
three azimuthal angles (0$^\circ$, 45$^\circ$ and 90$^\circ$). The polarized flux maps are presented in Fig.~\ref{Fig:Maps_TF},
the polarization fraction maps in Fig.~\ref{Fig:Maps_PO} and the polarization position angle maps in Fig.~\ref{Fig:Maps_PA}.
The spatial resolution of the images is 200$\times$200 pixels which, for a NGC~1068-like AGN distance would give a pixel size 
of 0.11$\times$0.11~pc (0.1~arcsec $\approx$ 11~$h_{50}^{-1}$~pc, see \citealt{Capetti1995}). The (0,0) coordinates give the position of the
central SMBH. 

As we can see from imaging polarimetry (Fig.~\ref{Fig:Maps_TF}, \ref{Fig:Maps_PO} and \ref{Fig:Maps_PA}), most of 
the polarized flux emerges from the circumnuclear dust funnel and propagates through the wind. Forward scattering inside the wind 
happens too rarely to impact the larger amount of flux that has scattered inside the inner dust radius or propagated into the 
torus then escaped along the polar direction. Equatorial scattering thus imposes its $PPA$, as it can been seen from 
Fig.~\ref{Fig:Maps_PA}. The net polarization degree is diluted by unpolarized radiation from the source, resulting in low values 
(Fig.~\ref{Fig:Maps_PO}). The elongated shape of the spot of maximum flux is due to the inclination of the system in 
addition to backward scattering of photons from the torus to the observer. Thanks to the gaps between the clouds, photons can journey 
further into the dust medium than in the case of a smooth-density distribution but their polarization signature has no global impact 
\citep{Marin2015}. The warps are not easily detected from imaging polarimetry of Seyfert-1s. For intermediate inclinations, 
the line-of-sight of the observer is not obscured by the equatorial dust. The inner region shines and its polarization dominates over 
the polarization resulting from wind interactions. This is a visual confirmation of the results presented in Fig.~\ref{Fig:Table_AGN_PO} 
and \ref{Fig:Table_AGN_PA}. The warp can be distinguished in the maps (middle-right images) as an extension towards the left of the 
image. This extension is particularly visible in the polarization degree and polarization angle maps while, due to the small 
amount of scattered radiation, can be missed in the polarized flux maps. Without prior information, it is impossible to determine if 
this material is an isolated dust lane or if it is part of the dusty structure. The warp that is situated beneath the equatorial plane 
is invisible. It is at Seyfert-2 inclinations that the warp has a direct impact onto the polarization maps. At an azimuthal 
angle of 0$^\circ$, when the warp's maximum polar elongation faces the observer, an important fraction of the wind base is obscured 
and there is an asymmetry in the polarized flux detected from the northern AGN part with respect to the southern flux. This is not 
due to an orientation effect, the warp is directly impacting the observed distribution of fluxes, with the southern cone being much 
brighter than its northern counterpart. This naturally explains why most of the variations in $P$ and $PPA$ observed in 
Fig.~\ref{Fig:Table_AGN_PO} and \ref{Fig:Table_AGN_PA} occurs at an azimuthal angle of 0$^\circ$. Rotating around the AGN slowly 
uncovers the northern wind base and equilibrates the distribution of polarized fluxes. When the maximum asymmetry of the model is 
reached (azimuthal angle 90$^\circ$), both the upper and lower warps can be detected thanks to multiple scattering, but their 
relative importance is too weak to impact the integrated polarization properties (see the low polarization fractions on 
Fig.~\ref{Fig:Maps_PO}). The ratio between the polar scattered flux and the flux from the warped torus surfaces tells 
us that warped tori might be difficult to detect even with current polarimetric observatories.

\section{Exploring the warp morphology}
\label{Jud}

\begin{figure}
   \centering
   \includegraphics[trim = 0mm 0mm 0mm 0mm, clip, width=9cm]{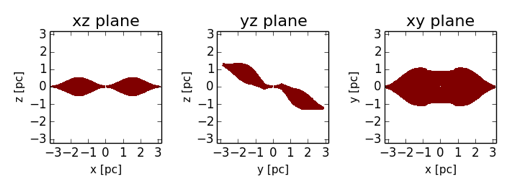}
   \caption{Cuts through the dust density distribution of the standard model shown in 
	    \citet{Jud2017}, taken along the coordinate planes.
	    The model was clumpy and then implemented in {\sc stokes}.}
  \label{Fig:Jud_sketch}
\end{figure}

We have seen that warped dusty tori are acceptable within the unified model picture as long as they do not violate one of the most important 
rules of the paradigm: there must be a polarization dichotomy between Seyfert-1 and Seyfert-2 orientations. Due to the clumpiness of the 
circumnuclear structure and the parametrization of the warps, a limited number of exceptions can arise but the majority of the models 
have proven to pass the test when adding additional AGN components. Yet, the question of the importance of the warp morphology remains 
as the methodology to produce warps is not unique. To illustrate how a similar model can change the resulting polarization degree and 
angle, we took the dust distribution from \citet{Jud2017} and implemented it into {\sc stokes}. We clumpy the structure such that 
its volume filling factor is the same as the models investigated so far. A sketch of the corresponding smooth-density distribution model is shown in 
Fig.~\ref{Fig:Jud_sketch}.

\begin{figure}
   \centering
   \includegraphics[trim = 0mm 0mm 0mm 0mm, clip, width=9.2cm]{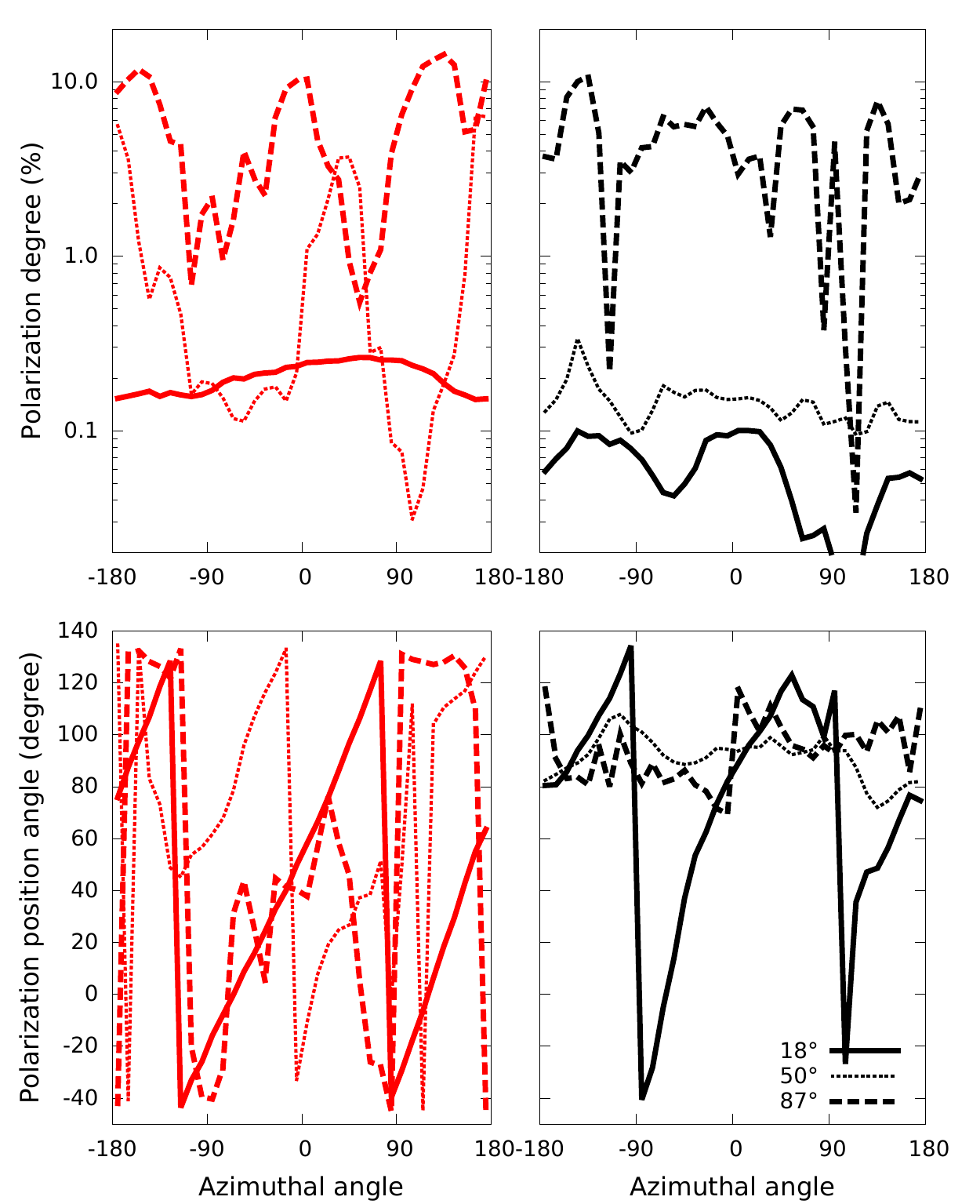}
   \caption{Left panels: optical polarization (5000~\AA) from the warped dusty 
	    disk presented in \citet{Jud2017}. For comparison, a warped 
	    torus created with the formulas presented in this paper using 
	    R$_{\rm warp}$ = 0.5~pc and $\theta_{\rm warp}$ = 10$^\circ$
	    is shown (black, right panels). The parametrization of Jud's disk 
	    is the same as in their paper and the model was clumpy in 
	    order to achieve a 22--23\% volume filling factor.}
  \label{Fig:Jud_disk}
\end{figure}

Running our Monte Carlo code for an isolated, warped and clumpy dusty torus, we found several similarities and differences with respect 
to the baseline model presented in this paper. In Fig.~\ref{Fig:Jud_disk}, we present the polarization results for the Jud model (left hand 
side panels, in red) and for a warped torus (right hand side panels, in black) created with the formulas presented in this paper. We used 
R$_{\rm warp}$ = 0.5~pc and $\theta_{\rm warp}$ = 10$^\circ$ to be as close as possible to the former structure. It appears that the model 
from \citet{Jud2017} produces different amounts of polarization degree associated with sharp sawtooth variations of the $PPA$ when the observer 
is revolving around the structure. The polarization angle modulations happen at the same azimuthal angle, with minor variations due to the random
distribution of clumps. However, the polarization degree in Jud models is always superior to 0.1\%, while in the case of the warped tori 
discussed in the previous sections of this paper, $P$ was often lower than 0.1\%. The sharp diminution in $P$ associated with the $PPA$ rotations 
are not seen here due to the puffed-up structure of the warped component that enhances the scattering probabilities. This peculiarity could be 
used to constrain the geometry of warped tori around point sources\footnote{Such as warps in proto-planetary and dusty debris disks 
\citep{Heap2000,Schneider2014,Marino2015}.}. In the cases of the intermediate and equatorial viewing angles, the observer's line-of-sight is 
passing through the bulk of the obscuring material but the deep $PPA$ polarization modulations still occur. Compared to the models of this paper, the geometry 
of the warped disk from \citet{Jud2017} presents high and rapid variations of the polarization degree and no stabilization of the polarization 
angle around 0$^\circ$. 

\begin{figure}
   \centering
   \includegraphics[trim = 0mm 0mm 0mm 0mm, clip, width=9.2cm]{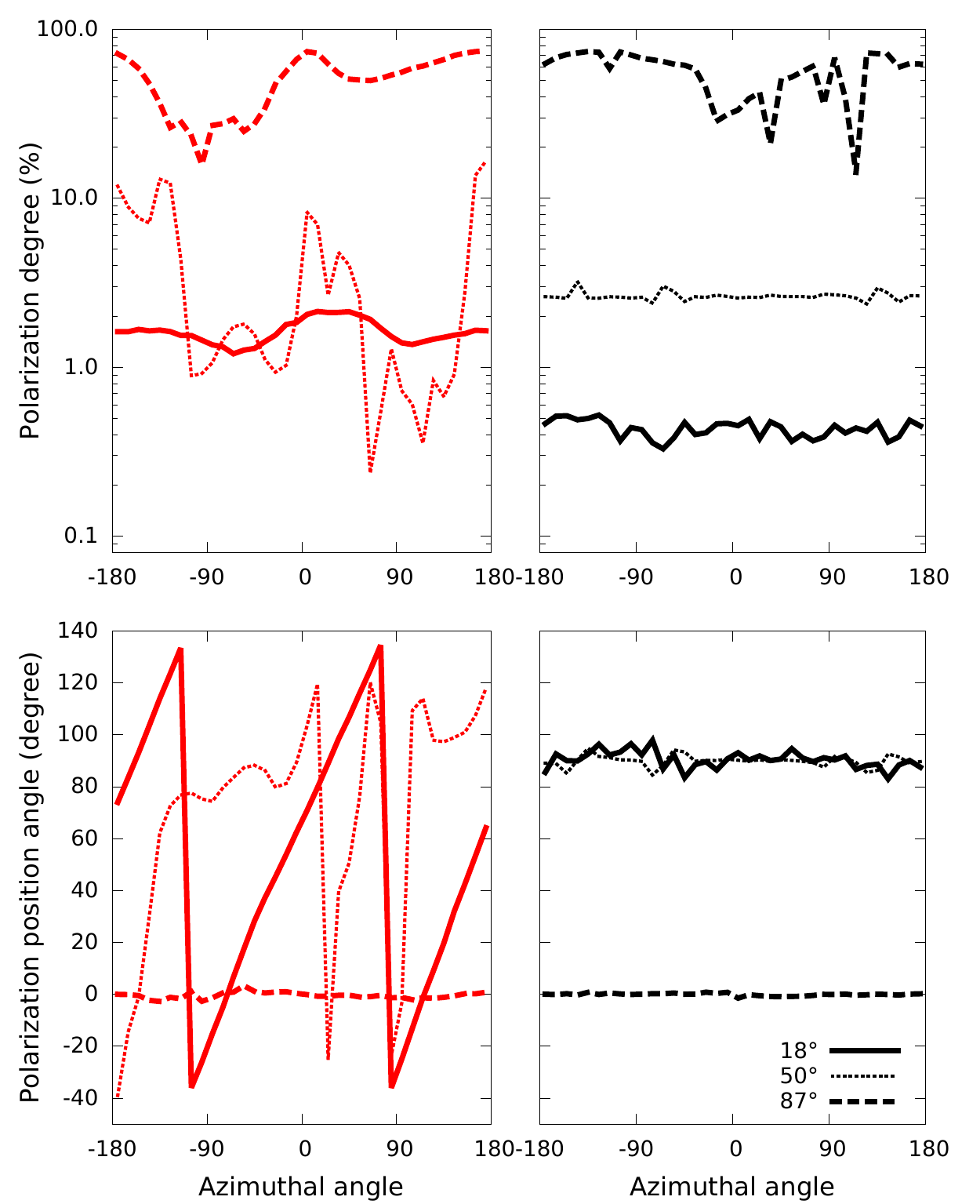}
   \caption{Same as Fig.~\ref{Fig:Jud_disk}, with 
	    the addition of a radiation-supported 
	    disk and polar outflows. The final models 
	    are thus similar to the AGN prescription
	    studied in Sect.~\ref{AGN}.}
  \label{Fig:Jud_AGN}
\end{figure}

Including the other two components (equatorial radiation-supported disk and ionization cones) to the warped torus allows us to check 
the polarization properties of an hypothetical AGN with a Jud's dusty disk (Fig.~\ref{Fig:Jud_AGN}, left hand side panels, in red), 
compared to our warped toy-model (Fig.~\ref{Fig:Jud_AGN}, right hand side panels, in black). The properties of the extra components are 
the same as before so differences in the AGN polarization signatures will reflect the impact of the warped structure. The most striking 
result of Fig.~\ref{Fig:Jud_AGN} is that the model violates the predictions from the unified model: for face-on and intermediate inclinations, 
the polarization position angle is neither parallel nor perpendicular to the projected polar axis of the system but has preserved its 
sawtooth oscillations despite the additional scattering components. The difference is due to the geometry of the torus: 1) the model from 
\citet{Jud2017} has a concave shape (e.g. a sphere cut out) whereas the warped torus model from this paper has a convex shape in the 
innermost region (an elliptical cross section, \citealt{Goosmann2007}), and 2) the width of the warped dusty medium from \citet{Jud2017},
extending towards the polar region, is twice as large as in the model from this paper. The change in morphology enhances the scattering 
probabilities for the Jud model and more scattered radiation bounces from the dust structure beyond R$_{\rm warp}$ to a polar observer. 
The polarization signatures from the warped torus are preserved and result in variable $PPA$. Together with the degree of polarization 
(1 to 2\%), this is unfortunately in contradiction to observations. Those high degrees of $P$ are rare among Seyfert-1s and are most of the 
time associated with a perpendicular polarization angle (the so-called polar-scattering dominated AGN, see \citealt{Smith2002} and 
\citealt{Marin2016b}). According to this model, all pole-on AGN should present high degrees of polarization together with an apparent random 
distribution of $PPA$. The case is similar for intermediate inclinations but it is more aligned with the prediction of the AGN paradigm. 
Intermediate orientation are allowed to produce $P >$ 1\% and the simulation shows that the polarization angle is plateauing at 0$^\circ$ 
or 90$^\circ$ according to the phase. For edge-on inclinations, the model is dominated by scattering inside the electron cone and is in 
agreement with polarimetric expectations of Seyferts ($P$ = 20--80\% and perpendicular polarization angle, see \citealt{Marin2014b}).

\section{Outlook}
\label{Discussion}
In this paper, we presented an optical investigation of the continuum polarization from warped tori. The time and resources required by 
the simulations to run polarized transfer in a clumpy medium are high enough to explain why such work was never undertaken previously. 
With the development of faster codes, a more complete view of the polarimetric signature of warped and clumpy media is in reach,
especially extending the wavelength range of our work. In particular, two codes are perfectly suited in the infrared band. {\sc skirt}, 
presented in \citet{Baes2003}, \citet{Baes2011} and \citet{Camps2015}, is a radiative transfer code that simulates the journey of photon 
packages in dusty astrophysical systems. Thanks to an adaptive grid, {\sc skirt} handles any 3D geometry of the dust distribution and 
offers a full treatment of multiple anisotropic scattering, absorbing and (re-)emitting media, which makes it a perfect tool to extend
the work accomplished in this paper. The polarization of radiation by spherical dust grains is included and soon should be released 
to the public (Peest et al, submitted). The second infrared tool, MontAGN, is dedicated to the production of high angular resolution 
polarization maps (Grosset et al., in prep). The code shares many common points with {\sc stokes} and also includes re-emission by dust, 
computing temperature equilibrium adjustment at each absorption. MontAGN works by sending photon packets that travel in a user-defined grid 
of dusty cells, such as {\sc skirt}, and overlaps the wavelengths band of {\sc stokes} in the 0.8 -- 1~$\mu$m regime. This allowed for a 
comparison between the codes \citep{Grosset2016,Marin2016c}, which gave satisfactory results. The use of a grid in both infrared codes makes 
the creation of a clumpy medium easy and computationally cheap. Looking at the infrared polarization, it might become easier to check whether 
a defined warped configuration can explain the observed infrared polarization of nearby Seyferts, intrinsically less polluted by starlight 
emission than in the optical band. The importance of re-emission and multiple scattering, depolarizing the escaping radiation, will 
probably be enhanced by clumpiness and may lead to different signatures in $PPA$ than in the V and B bands.

Another waveband of importance is the X-ray band. With the advance of modern polarimeters, either based on the photoelectric effect 
\citep[e.g.,][]{Costa2001}, gas-filled Time Projection Chambers \citep[e.g.,][]{Black2007}, or Compton scattering \citep[e.g.,][]{Guo2013},
it will be soon possible to measure the X-ray polarization from nearby and bright AGN. The expected AGN polarization in the soft energy band 
(2 -- 10~keV) should be of similar levels than in the optical band \citep{Marin2016d} but due to Compton and Inverse-Compton scattering, 
the hard X-ray band might show a distinctive polarization. It is thus important to produce new simulations of AGN including clumpy 
tori based on contemporary observations \citep{Marinucci2016}. Preliminary results for smooth density warped disks were recently obtained 
by \citet{Cheng2016}, who computed the X-ray polarimetric features induced by a warped disk around a stellar-mass black hole. Using 
the Bardeen-Petterson effect, the authors explored the possibility of identifying a warped disk with X-ray spectropolarimetric measurements 
of the thermal spectrum of the accretion disk. Among other findings, they showed that the warped structure can be identified by a transition
in polarization in the 0.1 -- 3~keV band. The larger the inclination between the inner and outer part of the warped accretion disk, 
the more prominent is the feature. By looking from a polar direction, it might be possible to detect this feature with a future mission
but the surrounding of the accreting system must be taken into account first (polar outflows being among the most important 
regions to add). The detection of an X-ray polarization signal influenced by a warped accreting medium might even be enhanced by 
Lense--Thirring precession of the inner accretion flow. \citet{Ingram2015} found that the polarization signature emitted from a
truncated disk/precessing inner flow geometry in an X-ray binary oscillates at a specific frequency where quasi-periodic oscillations 
are detected. The maximum polarization degree due to the Lense--Thirring effect is of the order of 1.5~\%, which is not negligible for 
a potential detection. The case for AGN remains to be explored but encouraging results are emerging for the detection and characterization 
of warped structures around compact objects.

\section{Conclusions}
\label{Conclusion}
In this paper, we explored the optical, scattering-induced, continuum polarization emerging from clumpy and warped dusty tori. 
Significant features appear at Seyfert-1 inclinations, where the polarization position angle rotates by $\sim$~180$^\circ$ in sawtooth oscillations, 
locally decreasing the degree of polarization. At intermediate and edge-on inclinations, both $P$ and the $PPA$ are similar to non-warped 
clumpy tori. Examining a large range of warp parameters, we found that the impact of clumpiness onto $P$ is overwhelming the signatures of 
warps and it is almost impossible to distinguish between two different warped structures. 

Including a warped torus in a more complex AGN model, with equatorial inflows and polar outflows, we are able to recover the observed 
polarization properties of nearby Seyferts, together with a proper distinction in polarization angles between a Seyfert-1 and a Seyfert-2. 
This indicates that warped equatorial structures may exist in AGN, leaving very little imprints on the optical continuum. However, 
imaging polarimetry is able to reveal the warps in Seyfert-2s but the sensitivity, contrast and angular resolution needed to achieve 
a detection are beyond current observatories, despite the progresses made by direct imaging instruments equipped with extreme adaptive 
optics systems \citep{Gratadour2015}.

If warped equatorial tori may indeed exist in AGN and pass the polarimetric tests of the unified model, the exact geometry of the warped 
structure must be examined carefully. We have proven that not all geometries are able to reproduce the observed polarization dichotomy. 
In this case, any new model of dusty disk-born winds \citep{Czerny2011}, warped disks \citep{Lawrence2010} or exotic self-gravitating 
tori \citep{Trova2016} must be investigated using polarimetry. Polarized radiative transfer simulations will then ensure that the 
structure is consistent with the most geometry-sensitive tool the unified model has to offer.

\acknowledgements 
The authors would like to acknowledge the anonymous referee for her/his useful comments that improved 
the quality of the paper. The authors are also grateful to D. Gratadour and L. Grosset from LESIA; 
M. Stalevski from the University of Chile; R. W. Goosmann and B. Vollmer from the Observatory of Strasbourg; 
K. Tristram from the European Southern Observatory, L. Burtscher from the faculty of Sterrewacht Leiden; and
R. Antonucci from the University of California for useful discussions and comments that greatly helped to 
ameliorate this manuscript.

\bibliographystyle{aa}
\bibliography{biblio}

\begin{thebibliography}{90}
\expandafter\ifx\csname natexlab\endcsname\relax\def\natexlab#1{#1}\fi

\bibitem[{{Antonucci}(1993)}]{Antonucci1993}
{Antonucci}, R. 1993, \araa, 31, 473

\bibitem[{{Antonucci}(1984)}]{Antonucci1984}
{Antonucci}, R.~R.~J. 1984, \apj, 278, 499

\bibitem[{{Antonucci} \& {Miller}(1985)}]{Antonucci1985}
{Antonucci}, R.~R.~J. \& {Miller}, J.~S. 1985, \apj, 297, 621

\bibitem[{{Asmus} {et~al.}(2016){Asmus}, {H{\"o}nig}, \& {Gandhi}}]{Asmus2016}
{Asmus}, D., {H{\"o}nig}, S.~F., \& {Gandhi}, P. 2016, \apj, 822, 109

\bibitem[{{Augereau} {et~al.}(2001){Augereau}, {Nelson}, {Lagrange},
  {Papaloizou}, \& {Mouillet}}]{Augereau2001}
{Augereau}, J.~C., {Nelson}, R.~P., {Lagrange}, A.~M., {Papaloizou}, J.~C.~B.,
  \& {Mouillet}, D. 2001, \aap, 370, 447

\bibitem[{{Baes} {et~al.}(2003){Baes}, {Davies}, {Dejonghe}, {Sabatini},
  {Roberts}, {Evans}, {Linder}, {Smith}, \& {de Blok}}]{Baes2003}
{Baes}, M., {Davies}, J.~I., {Dejonghe}, H., {et~al.} 2003, \mnras, 343, 1081

\bibitem[{{Baes} {et~al.}(2011){Baes}, {Verstappen}, {De Looze}, {Fritz},
  {Saftly}, {Vidal P{\'e}rez}, {Stalevski}, \& {Valcke}}]{Baes2011}
{Baes}, M., {Verstappen}, J., {De Looze}, I., {et~al.} 2011, \apjs, 196, 22

\bibitem[{{Bardeen} \& {Petterson}(1975)}]{Bardeen1975}
{Bardeen}, J.~M. \& {Petterson}, J.~A. 1975, \apjl, 195, L65

\bibitem[{{Black} {et~al.}(2007){Black}, {Baker}, {Deines-Jones}, {Hill}, \&
  {Jahoda}}]{Black2007}
{Black}, J.~K., {Baker}, R.~G., {Deines-Jones}, P., {Hill}, J.~E., \& {Jahoda},
  K. 2007, Nuclear Instruments and Methods in Physics Research A, 581, 755

\bibitem[{{Bock} {et~al.}(2000){Bock}, {Neugebauer}, {Matthews}, {Soifer},
  {Becklin}, {Ressler}, {Marsh}, {Werner}, {Egami}, \& {Blandford}}]{Bock2000}
{Bock}, J.~J., {Neugebauer}, G., {Matthews}, K., {et~al.} 2000, \aj, 120, 2904

\bibitem[{{Burtscher} {et~al.}(2009){Burtscher}, {Jaffe}, {Raban},
  {Meisenheimer}, {Tristram}, \& {R{\"o}ttgering}}]{Burtscher2009}
{Burtscher}, L., {Jaffe}, W., {Raban}, D., {et~al.} 2009, \apjl, 705, L53

\bibitem[{{Burtscher} {et~al.}(2013){Burtscher}, {Meisenheimer}, {Tristram},
  {Jaffe}, {H{\"o}nig}, {Davies}, {Kishimoto}, {Pott}, {R{\"o}ttgering},
  {Schartmann}, {Weigelt}, \& {Wolf}}]{Burtscher2013}
{Burtscher}, L., {Meisenheimer}, K., {Tristram}, K.~R.~W., {et~al.} 2013, \aap,
  558, A149

\bibitem[{{Camps} \& {Baes}(2015)}]{Camps2015}
{Camps}, P. \& {Baes}, M. 2015, Astronomy and Computing, 9, 20

\bibitem[{{Capetti} {et~al.}(1996){Capetti}, {Axon}, {Macchetto}, {Sparks}, \&
  {Boksenberg}}]{Capetti1996}
{Capetti}, A., {Axon}, D.~J., {Macchetto}, F., {Sparks}, W.~B., \&
  {Boksenberg}, A. 1996, \apj, 466, 169

\bibitem[{{Capetti} {et~al.}(1995){Capetti}, {Macchetto}, {Axon}, {Sparks}, \&
  {Boksenberg}}]{Capetti1995}
{Capetti}, A., {Macchetto}, F., {Axon}, D.~J., {Sparks}, W.~B., \&
  {Boksenberg}, A. 1995, \apjl, 452, L87

\bibitem[{{Caproni} \& {Abraham}(2004)}]{Caproni2004b}
{Caproni}, A. \& {Abraham}, Z. 2004, \apj, 602, 625

\bibitem[{{Caproni} {et~al.}(2004){Caproni}, {Mosquera Cuesta}, \&
  {Abraham}}]{Caproni2004a}
{Caproni}, A., {Mosquera Cuesta}, H.~J., \& {Abraham}, Z. 2004, \apjl, 616, L99

\bibitem[{{Cheng} {et~al.}(2016){Cheng}, {Liu}, {Nampalliwar}, \&
  {Bambi}}]{Cheng2016}
{Cheng}, Y., {Liu}, D., {Nampalliwar}, S., \& {Bambi}, C. 2016, Classical and
  Quantum Gravity, 33, 125015

\bibitem[{{Costa} {et~al.}(2001){Costa}, {Soffitta}, {Bellazzini}, {Brez},
  {Lumb}, \& {Spandre}}]{Costa2001}
{Costa}, E., {Soffitta}, P., {Bellazzini}, R., {et~al.} 2001, \nat, 411, 662

\bibitem[{{Czerny} \& {Hryniewicz}(2011)}]{Czerny2011}
{Czerny}, B. \& {Hryniewicz}, K. 2011, \aap, 525, L8

\bibitem[{{Gallimore} {et~al.}(2004){Gallimore}, {Baum}, \&
  {O'Dea}}]{Gallimore2004}
{Gallimore}, J.~F., {Baum}, S.~A., \& {O'Dea}, C.~P. 2004, \apj, 613, 794

\bibitem[{{Gillessen} {et~al.}(2006){Gillessen}, {Perrin}, {Brandner},
  {Straubmeier}, {Eisenhauer}, {Rabien}, {Eckart}, {Lena}, {Genzel}, {Paumard},
  \& {Hippler}}]{Gillessen2006}
{Gillessen}, S., {Perrin}, G., {Brandner}, W., {et~al.} 2006, in \procspie,
  Vol. 6268, Society of Photo-Optical Instrumentation Engineers (SPIE)
  Conference Series, 626811

\bibitem[{{Goosmann} \& {Gaskell}(2007)}]{Goosmann2007}
{Goosmann}, R.~W. \& {Gaskell}, C.~M. 2007, \aap, 465, 129

\bibitem[{{Gratadour} {et~al.}(2015){Gratadour}, {Rouan}, {Grosset},
  {Boccaletti}, \& {Cl{\'e}net}}]{Gratadour2015}
{Gratadour}, D., {Rouan}, D., {Grosset}, L., {Boccaletti}, A., \& {Cl{\'e}net},
  Y. 2015, \aap, 581, L8

\bibitem[{{Grosset} {et~al.}(2016){Grosset}, {Marin}, {Gratadour}, {Goosmann},
  {Rouan}, {Cl{\'e}net}, {Pelat}, \& {Rojas Lobos}}]{Grosset2016}
{Grosset}, L., {Marin}, F., {Gratadour}, D., {et~al.} 2016, ArXiv e-prints

\bibitem[{{Gunn}(1979)}]{Gunn1979}
{Gunn}, J.~E. 1979, {Feeding the monster - Gas discs in elliptical galaxies},
  ed. C.~{Hazard} \& S.~{Mitton}, 213--225

\bibitem[{{Guo} {et~al.}(2013){Guo}, {Beilicke}, {Garson}, {Kislat}, {Fleming},
  \& {Krawczynski}}]{Guo2013}
{Guo}, Q., {Beilicke}, M., {Garson}, A., {et~al.} 2013, Astroparticle Physics,
  41, 63

\bibitem[{{Hao} {et~al.}(2007){Hao}, {Weedman}, {Spoon}, {Marshall},
  {Levenson}, {Elitzur}, \& {Houck}}]{Hao2007}
{Hao}, L., {Weedman}, D.~W., {Spoon}, H.~W.~W., {et~al.} 2007, \apjl, 655, L77

\bibitem[{{Heap} {et~al.}(2000){Heap}, {Lindler}, {Lanz}, {Cornett}, {Hubeny},
  {Maran}, \& {Woodgate}}]{Heap2000}
{Heap}, S.~R., {Lindler}, D.~J., {Lanz}, T.~M., {et~al.} 2000, \apj, 539, 435

\bibitem[{{Herrnstein} {et~al.}(1999){Herrnstein}, {Moran}, {Greenhill},
  {Diamond}, {Inoue}, {Nakai}, {Miyoshi}, {Henkel}, \&
  {Riess}}]{Herrnstein1999}
{Herrnstein}, J.~R., {Moran}, J.~M., {Greenhill}, L.~J., {et~al.} 1999, \nat,
  400, 539

\bibitem[{{H{\"o}nig} {et~al.}(2006){H{\"o}nig}, {Beckert}, {Ohnaka}, \&
  {Weigelt}}]{Hoenig2006}
{H{\"o}nig}, S.~F., {Beckert}, T., {Ohnaka}, K., \& {Weigelt}, G. 2006, \aap,
  452, 459

\bibitem[{{Ingram} {et~al.}(2015){Ingram}, {Maccarone}, {Poutanen}, \&
  {Krawczynski}}]{Ingram2015}
{Ingram}, A., {Maccarone}, T.~J., {Poutanen}, J., \& {Krawczynski}, H. 2015,
  \apj, 807, 53

\bibitem[{{Jaffe} {et~al.}(2004){Jaffe}, {Meisenheimer}, {R{\"o}ttgering},
  {Leinert}, {Richichi}, {Chesneau}, {Fraix-Burnet}, {Glazenborg-Kluttig},
  {Granato}, {Graser}, {Heijligers}, {K{\"o}hler}, {Malbet}, {Miley},
  {Paresce}, {Pel}, {Perrin}, {Przygodda}, {Schoeller}, {Sol}, {Waters},
  {Weigelt}, {Woillez}, \& {de Zeeuw}}]{Jaffe2004}
{Jaffe}, W., {Meisenheimer}, K., {R{\"o}ttgering}, H.~J.~A., {et~al.} 2004,
  \nat, 429, 47

\bibitem[{{Jud} {et~al.}(2017){Jud}, {Schartmann}, {Mould}, {Burtscher}, \&
  {Tristram}}]{Jud2017}
{Jud}, H., {Schartmann}, M., {Mould}, J., {Burtscher}, L., \& {Tristram},
  K.~R.~W. 2017, \mnras, 465, 248

\bibitem[{{Kartje}(1995)}]{Kartje1995}
{Kartje}, J.~F. 1995, \apj, 452, 565

\bibitem[{{Kishimoto} {et~al.}(2001){Kishimoto}, {Antonucci}, {Cimatti},
  {Hurt}, {Dey}, {van Breugel}, \& {Spinrad}}]{Kishimoto2001}
{Kishimoto}, M., {Antonucci}, R., {Cimatti}, A., {et~al.} 2001, \apj, 547, 667

\bibitem[{{Kishimoto} {et~al.}(2009{\natexlab{a}}){Kishimoto}, {H{\"o}nig},
  {Antonucci}, {Kotani}, {Barvainis}, {Tristram}, \&
  {Weigelt}}]{Kishimoto2009a}
{Kishimoto}, M., {H{\"o}nig}, S.~F., {Antonucci}, R., {et~al.}
  2009{\natexlab{a}}, \aap, 507, L57

\bibitem[{{Kishimoto} {et~al.}(2007){Kishimoto}, {H{\"o}nig}, {Beckert}, \&
  {Weigelt}}]{Kishimoto2007}
{Kishimoto}, M., {H{\"o}nig}, S.~F., {Beckert}, T., \& {Weigelt}, G. 2007,
  \aap, 476, 713

\bibitem[{{Kishimoto} {et~al.}(2009{\natexlab{b}}){Kishimoto}, {H{\"o}nig},
  {Tristram}, \& {Weigelt}}]{Kishimoto2009}
{Kishimoto}, M., {H{\"o}nig}, S.~F., {Tristram}, K.~R.~W., \& {Weigelt}, G.
  2009{\natexlab{b}}, \aap, 493, L57

\bibitem[{{Kiuchi} {et~al.}(2015){Kiuchi}, {Sekiguchi}, {Kyutoku}, {Shibata},
  {Taniguchi}, \& {Wada}}]{Kiuchi2015}
{Kiuchi}, K., {Sekiguchi}, Y., {Kyutoku}, K., {et~al.} 2015, \prd, 92, 064034

\bibitem[{{Kollmeier} {et~al.}(2006){Kollmeier}, {Onken}, {Kochanek}, {Gould},
  {Weinberg}, {Dietrich}, {Cool}, {Dey}, {Eisenstein}, {Jannuzi}, {Le Floc'h},
  \& {Stern}}]{Kollmeier2006}
{Kollmeier}, J.~A., {Onken}, C.~A., {Kochanek}, C.~S., {et~al.} 2006, \apj,
  648, 128

\bibitem[{{Krolik} \& {Begelman}(1988)}]{Krolik1988}
{Krolik}, J.~H. \& {Begelman}, M.~C. 1988, \apj, 329, 702

\bibitem[{{Lawrence} \& {Elvis}(2010)}]{Lawrence2010}
{Lawrence}, A. \& {Elvis}, M. 2010, \apj, 714, 561

\bibitem[{{L{\'o}pez-Gonzaga} {et~al.}(2016){L{\'o}pez-Gonzaga}, {Burtscher},
  {Tristram}, {Meisenheimer}, \& {Schartmann}}]{Lopez2016}
{L{\'o}pez-Gonzaga}, N., {Burtscher}, L., {Tristram}, K.~R.~W., {Meisenheimer},
  K., \& {Schartmann}, M. 2016, \aap, 591, A47

\bibitem[{{Maiolino} {et~al.}(2010){Maiolino}, {Risaliti}, {Salvati},
  {Pietrini}, {Torricelli-Ciamponi}, {Elvis}, {Fabbiano}, {Braito}, \&
  {Reeves}}]{Maiolino2010}
{Maiolino}, R., {Risaliti}, G., {Salvati}, M., {et~al.} 2010, \aap, 517, A47

\bibitem[{{Manske} {et~al.}(1998){Manske}, {Henning}, \&
  {Men'shchikov}}]{Manske1998}
{Manske}, V., {Henning}, T., \& {Men'shchikov}, A.~B. 1998, \aap, 331, 52

\bibitem[{{Marin}(2014)}]{Marin2014b}
{Marin}, F. 2014, \mnras, 441, 551

\bibitem[{{Marin}(2016)}]{Marin2016b}
{Marin}, F. 2016, \mnras, 460, 3679

\bibitem[{{Marin} \& {Antonucci}(2016)}]{Marin2016}
{Marin}, F. \& {Antonucci}, R. 2016, \apj, 830, 82

\bibitem[{{Marin} \& {Goosmann}(2014)}]{Marin2014}
{Marin}, F. \& {Goosmann}, R.~W. 2014, in SF2A-2014: Proceedings of the Annual
  meeting of the French Society of Astronomy and Astrophysics, ed. J.~{Ballet},
  F.~{Martins}, F.~{Bournaud}, R.~{Monier}, \& C.~{Reyl{\'e}}, 103--108

\bibitem[{{Marin} {et~al.}(2015){Marin}, {Goosmann}, \& {Gaskell}}]{Marin2015}
{Marin}, F., {Goosmann}, R.~W., \& {Gaskell}, C.~M. 2015, \aap, 577, A66

\bibitem[{{Marin} {et~al.}(2012){Marin}, {Goosmann}, {Gaskell}, {Porquet}, \&
  {Dov{\v c}iak}}]{Marin2012}
{Marin}, F., {Goosmann}, R.~W., {Gaskell}, C.~M., {Porquet}, D., \& {Dov{\v
  c}iak}, M. 2012, \aap, 548, A121

\bibitem[{{Marin} {et~al.}(2016{\natexlab{a}}){Marin}, {Goosmann}, \&
  {Petrucci}}]{Marin2016d}
{Marin}, F., {Goosmann}, R.~W., \& {Petrucci}, P.-O. 2016{\natexlab{a}}, \aap,
  591, A23

\bibitem[{{Marin} {et~al.}(2016{\natexlab{b}}){Marin}, {Grosset}, {Goosmann},
  {Gratadour}, {Rouan}, {Cl{\'e}net}, {Pelat}, \& {Rojas Lobos}}]{Marin2016c}
{Marin}, F., {Grosset}, L., {Goosmann}, R., {et~al.} 2016{\natexlab{b}}, ArXiv
  e-prints

\bibitem[{{Marin} \& {Stalevski}(2015)}]{Marin2015b}
{Marin}, F. \& {Stalevski}, M. 2015, in SF2A-2015: Proceedings of the Annual
  meeting of the French Society of Astronomy and Astrophysics, ed.
  F.~{Martins}, S.~{Boissier}, V.~{Buat}, L.~{Cambr{\'e}sy}, \& P.~{Petit},
  167--170

\bibitem[{{Marino} {et~al.}(2015){Marino}, {Perez}, \& {Casassus}}]{Marino2015}
{Marino}, S., {Perez}, S., \& {Casassus}, S. 2015, \apjl, 798, L44

\bibitem[{{Marinucci} {et~al.}(2016){Marinucci}, {Bianchi}, {Matt},
  {Alexander}, {Balokovi{\'c}}, {Bauer}, {Brandt}, {Gandhi}, {Guainazzi},
  {Harrison}, {Iwasawa}, {Koss}, {Madsen}, {Nicastro}, {Puccetti}, {Ricci},
  {Stern}, \& {Walton}}]{Marinucci2016}
{Marinucci}, A., {Bianchi}, S., {Matt}, G., {et~al.} 2016, \mnras, 456, L94

\bibitem[{{Mason} {et~al.}(2009){Mason}, {Levenson}, {Shi}, {Packham},
  {Gorjian}, {Cleary}, {Rhee}, \& {Werner}}]{Mason2009}
{Mason}, R.~E., {Levenson}, N.~A., {Shi}, Y., {et~al.} 2009, \apjl, 693, L136

\bibitem[{{Mathis} {et~al.}(1977){Mathis}, {Rumpl}, \&
  {Nordsieck}}]{Mathis1977}
{Mathis}, J.~S., {Rumpl}, W., \& {Nordsieck}, K.~H. 1977, \apj, 217, 425

\bibitem[{{Menou} \& {Quataert}(2001)}]{Menou2001}
{Menou}, K. \& {Quataert}, E. 2001, \apj, 552, 204

\bibitem[{{Meyer}(1989)}]{Meyer1989}
{Meyer}, F., ed. 1989, NATO Advanced Science Institutes (ASI) Series C, Vol.
  290, {Theory of accretion disks}

\bibitem[{{Miller} {et~al.}(1991){Miller}, {Goodrich}, \&
  {Mathews}}]{Miller1991}
{Miller}, J.~S., {Goodrich}, R.~W., \& {Mathews}, W.~G. 1991, \apj, 378, 47

\bibitem[{{Nayakshin}(2005)}]{Nayakshin2005}
{Nayakshin}, S. 2005, \mnras, 359, 545

\bibitem[{{Nenkova} {et~al.}(2002){Nenkova}, {Ivezi{\'c}}, \&
  {Elitzur}}]{Nenkova2002}
{Nenkova}, M., {Ivezi{\'c}}, {\v Z}., \& {Elitzur}, M. 2002, \apjl, 570, L9

\bibitem[{{Nenkova} {et~al.}(2008{\natexlab{a}}){Nenkova}, {Sirocky},
  {Ivezi{\'c}}, \& {Elitzur}}]{Nenkova2008a}
{Nenkova}, M., {Sirocky}, M.~M., {Ivezi{\'c}}, {\v Z}., \& {Elitzur}, M.
  2008{\natexlab{a}}, \apj, 685, 147

\bibitem[{{Nenkova} {et~al.}(2008{\natexlab{b}}){Nenkova}, {Sirocky},
  {Nikutta}, {Ivezi{\'c}}, \& {Elitzur}}]{Nenkova2008b}
{Nenkova}, M., {Sirocky}, M.~M., {Nikutta}, R., {Ivezi{\'c}}, {\v Z}., \&
  {Elitzur}, M. 2008{\natexlab{b}}, \apj, 685, 160

\bibitem[{{Netzer}(2013)}]{Netzer2013}
{Netzer}, H. 2013, {The Physics and Evolution of Active Galactic Nuclei}

\bibitem[{{Nikutta} {et~al.}(2009){Nikutta}, {Elitzur}, \&
  {Lacy}}]{Nikutta2009}
{Nikutta}, R., {Elitzur}, M., \& {Lacy}, M. 2009, \apj, 707, 1550

\bibitem[{{Phinney}(1989)}]{Phinney1989}
{Phinney}, E.~S. 1989, in NATO Advanced Science Institutes (ASI) Series C, Vol.
  290, NATO Advanced Science Institutes (ASI) Series C, ed. F.~{Meyer}, 457

\bibitem[{{Pier} \& {Krolik}(1992)}]{Pier1992}
{Pier}, E.~A. \& {Krolik}, J.~H. 1992, \apj, 401, 99

\bibitem[{{Prevot} {et~al.}(1984){Prevot}, {Lequeux}, {Prevot}, {Maurice}, \&
  {Rocca-Volmerange}}]{Prevot1984}
{Prevot}, M.~L., {Lequeux}, J., {Prevot}, L., {Maurice}, E., \&
  {Rocca-Volmerange}, B. 1984, \aap, 132, 389

\bibitem[{{Pringle}(1996)}]{Pringle1996}
{Pringle}, J.~E. 1996, \mnras, 281, 357

\bibitem[{{Sazonov} {et~al.}(2015){Sazonov}, {Churazov}, \&
  {Krivonos}}]{Sazonov2015}
{Sazonov}, S., {Churazov}, E., \& {Krivonos}, R. 2015, \mnras, 454, 1202

\bibitem[{{Schartmann} {et~al.}(2008){Schartmann}, {Meisenheimer}, {Camenzind},
  {Wolf}, {Tristram}, \& {Henning}}]{Schartmann2008}
{Schartmann}, M., {Meisenheimer}, K., {Camenzind}, M., {et~al.} 2008, \aap,
  482, 67

\bibitem[{{Schneider} {et~al.}(2014){Schneider}, {Grady}, {Hines}, {Stark},
  {Debes}, {Carson}, {Kuchner}, {Perrin}, {Weinberger}, {Wisniewski},
  {Silverstone}, {Jang-Condell}, {Henning}, {Woodgate}, {Serabyn},
  {Moro-Martin}, {Tamura}, {Hinz}, \& {Rodigas}}]{Schneider2014}
{Schneider}, G., {Grady}, C.~A., {Hines}, D.~C., {et~al.} 2014, \aj, 148, 59

\bibitem[{{Schnorr-M{\"u}ller} {et~al.}(2016){Schnorr-M{\"u}ller}, {Davies},
  {Korista}, {Burtscher}, {Rosario}, {Storchi-Bergmann}, {Contursi}, {Genzel},
  {Graci{\'a}-Carpio}, {Hicks}, {Janssen}, {Koss}, {Lin}, {Lutz},
  {Maciejewski}, {M{\"u}ller-S{\'a}nchez}, {Orban de Xivry}, {Riffel},
  {Riffel}, {Schartmann}, {Sternberg}, {Sturm}, {Tacconi}, {Veilleux}, \&
  {Ulrich}}]{Schnorr2016}
{Schnorr-M{\"u}ller}, A., {Davies}, R.~I., {Korista}, K.~T., {et~al.} 2016,
  \mnras, 462, 3570

\bibitem[{{Shuder}(1981)}]{Shuder1981}
{Shuder}, J.~M. 1981, \apj, 244, 12

\bibitem[{{Siebenmorgen} {et~al.}(2015){Siebenmorgen}, {Heymann}, \&
  {Efstathiou}}]{Siebenmorgen2015}
{Siebenmorgen}, R., {Heymann}, F., \& {Efstathiou}, A. 2015, \aap, 583, A120

\bibitem[{{Smith} {et~al.}(2002){Smith}, {Young}, {Robinson}, {Corbett},
  {Giannuzzo}, {Axon}, \& {Hough}}]{Smith2002}
{Smith}, J.~E., {Young}, S., {Robinson}, A., {et~al.} 2002, \mnras, 335, 773

\bibitem[{{Stalevski} {et~al.}(2012){Stalevski}, {Fritz}, {Baes}, {Nakos}, \&
  {Popovi{\'c}}}]{Stalevski2012}
{Stalevski}, M., {Fritz}, J., {Baes}, M., {Nakos}, T., \& {Popovi{\'c}}, L.~{\v
  C}. 2012, \mnras, 420, 2756

\bibitem[{{Stokes}(1851)}]{Stokes1851}
{Stokes}, G.~G. 1851, Transactions of the Cambridge Philosophical Society, 9,
  399

\bibitem[{{Sturm} {et~al.}(2006){Sturm}, {Hasinger}, {Lehmann}, {Mainieri},
  {Genzel}, {Lehnert}, {Lutz}, \& {Tacconi}}]{Sturm2006}
{Sturm}, E., {Hasinger}, G., {Lehmann}, I., {et~al.} 2006, \apj, 642, 81

\bibitem[{{Swain} {et~al.}(2003){Swain}, {Vasisht}, {Akeson}, {Monnier},
  {Millan-Gabet}, {Serabyn}, {Creech-Eakman}, {van Belle}, {Beletic},
  {Beichman}, {Boden}, {Booth}, {Colavita}, {Gathright}, {Hrynevych},
  {Koresko}, {Le Mignant}, {Ligon}, {Mennesson}, {Neyman}, {Sargent}, {Shao},
  {Thompson}, {Unwin}, \& {Wizinowich}}]{Swain2003}
{Swain}, M., {Vasisht}, G., {Akeson}, R., {et~al.} 2003, \apjl, 596, L163

\bibitem[{{Tristram} {et~al.}(2014){Tristram}, {Burtscher}, {Jaffe},
  {Meisenheimer}, {H{\"o}nig}, {Kishimoto}, {Schartmann}, \&
  {Weigelt}}]{Tristram2014}
{Tristram}, K.~R.~W., {Burtscher}, L., {Jaffe}, W., {et~al.} 2014, \aap, 563,
  A82

\bibitem[{{Tristram} {et~al.}(2007){Tristram}, {Meisenheimer}, {Jaffe},
  {Schartmann}, {Rix}, {Leinert}, {Morel}, {Wittkowski}, {R{\"o}ttgering},
  {Perrin}, {Lopez}, {Raban}, {Cotton}, {Graser}, {Paresce}, \&
  {Henning}}]{Tristram2007}
{Tristram}, K.~R.~W., {Meisenheimer}, K., {Jaffe}, W., {et~al.} 2007, \aap,
  474, 837

\bibitem[{{Trova} {et~al.}(2016){Trova}, {Karas}, {Slan{\'y}}, \& {Kov{\'a}{\v
  r}}}]{Trova2016}
{Trova}, A., {Karas}, V., {Slan{\'y}}, P., \& {Kov{\'a}{\v r}}, J. 2016, \apjs,
  226, 12

\bibitem[{{Urry} \& {Padovani}(1995)}]{Urry1995}
{Urry}, C.~M. \& {Padovani}, P. 1995, \pasp, 107, 803

\bibitem[{{Wada} {et~al.}(2016){Wada}, {Schartmann}, \& {Meijerink}}]{Wada2016}
{Wada}, K., {Schartmann}, M., \& {Meijerink}, R. 2016, \apjl, 828, L19

\bibitem[{{Wolf} \& {Henning}(1999)}]{Wolf1999}
{Wolf}, S. \& {Henning}, T. 1999, \aap, 341, 675

\bibitem[{{Young}(2000)}]{Young2000}
{Young}, S. 2000, \mnras, 312, 567

\end{thebibliography}

\end{document}